\documentclass[a4,12pt,eclepsf]{article}
\usepackage{graphicx}
\usepackage[labelfont=bf]{caption} 
\usepackage{wrapfig, fancybox, fancyhdr, physics, amsmath, amssymb, makeidx, comment, multirow, bm, here, float}
\usepackage[colorlinks=true, linkcolor=blue, citecolor=blue, urlcolor=blue]{hyperref}

\newcommand{\BM}{\begin{minipage}}
\newcommand{\EM}{\end{minipage}}

\begin{document}

\renewcommand{\theequation}
{\thesection.\arabic{equation}}
\thispagestyle{empty}
\vspace*{5mm} 

\begin{center}
{\LARGE {\bf Holographic Subregion Complexity}} \\[5mm]
{\LARGE {\bf  and Fidelity Susceptibility}} \\[5mm]
{\LARGE {\bf  in Noncommutative Yang--Mills Theory}} 

\vspace*{15mm}

\renewcommand{\thefootnote}{\fnsymbol{footnote}}
{\Large Tadahito NAKAJIMA \footnote{email: nakajima.tadahito@nihon-u.ac.jp}} 
\vspace*{10mm}

{\it College of Engineering, Nihon University, Fukushima 963-8642, Japan} \\[15mm]

{\bf Abstract} \\[10mm]

\end{center}

We analyze the behavior of holographic subregion complexity (HSC) and holographic fidelity susceptibility (HFS) in noncommutative Yang--Mills theory. The emergence of a minimum length scale, dictated by the degree of noncommutativity, induces a behavioral transition in the HSC and establishes a lower bound. In the large noncommutativity regime, the qualitative features of the complexity deviate significantly from the commutative case. The HFS is shown to provide an effective measure of the degree of noncommutativity. Although the HSC generally satisfies strong subadditivity, this property fails abruptly when the subregion size approaches the minimum length scale. At finite temperature, the long-range behavior of the HSC is modified, and its lower bound scales positively with temperature. Furthermore, temperature enhances the sensitivity of the fidelity susceptibility to the degree of noncommutativity. Within the AdS soliton background, a competition between connected and disconnected configurations arises in the HSC, signaling a phase-transition-like behavior. Finally, the compactification scale is found to diminish the sensitivity of the HFS to the degree of noncommutativity. 

\clearpage

\setcounter{section}{0}
\section{Introduction}
\setcounter{page}{1}
\setcounter{equation}{0}

In recent years, complexity has come to be regarded as a fundamental quantity in quantum physics, standing alongside entanglement in its significance. Complexity is defined as the minimum number of gates required to prepare a quantum state from a reference state using a sequence of unitary gates acting on a small number of qubits \cite{LS2,MAN}. This concept has emerged as a key to understanding the growth of the black hole interior (the wormhole volume) that follows from the black hole geometry. Specifically, the complexity grows linearly with time until it reaches its maximum value \cite{ARBLS}.

The linear growth of the wormhole volume is conjectured to correspond to the linear growth of quantum complexity. This correspondence is expressed as the ``Complexity~${\cal C}$ = Volume~$V$'' (CV)  conjecture \cite{LS1, DSLS1}:
\begin{align}
\label{101}
{\cal C} =\dfrac{V}{G_{N}l} ,,
\end{align}
where $G_{N}$ and $l$ denote the Newton constant and the characteristic length scale (often associated with the AdS radius), respectively. Following this, it was proposed that the complexity of a quantum state is equal to the gravitational action of a spacetime region called the Wheeler--DeWitt patch; this is known as the ``Complexity~${\cal C}$ = Action~$A$'' (CA)  conjecture \cite{ARBDARLSBSYZ1, ARBDARLSBSYZ2}.

Recently, the CV conjecture was extended to define a holographic complexity for subsystems \cite{MA}. In this framework, analogous to how holographic entanglement entropy is determined by the area of the Ryu--Takayanagi surface \cite{SRTT1, SRTT2}, the subregion complexity is identified with the volume of the codimension-one bulk hypersurface enclosed by that surface. This quantity, termed ``holographic subregion complexity'' (HSC), has been extensively studied in various contexts \cite{OBADC,PRTS, ABABSM, SKJYZ}.

Reference \cite{MA} explores the connection between holographic subregion complexity and fidelity susceptibility. While fidelity serves as a measure of the overlap between quantum states, fidelity susceptibility characterizes the response of a state to infinitesimal changes in an external parameter. Notably, it has been shown that the leading term of the holographic subregion complexity, which follows a volume law, is qualitatively consistent with previous results regarding holographic fidelity susceptibility (HFS) \cite{MMTNNSTTKW}.

An interesting question arises as to how nonlocal interactions modify the properties of complexity and fidelity susceptibility. A prototypical field theory exhibiting nonlocal interactions is noncommutative Yang--Mills (NCYM) theory \cite{ACMRDAS, MRDCH, FAHASMMSJ, NSEW}. The noncommutativity of spacetime can induce nontrivial effects on a wide range of physical quantities \cite{MRS, AA, taka_naka-suzu, NST, TN_YO_KS, TN_YO_KS2, TN_YO_KS3, TN8}. In field theories with nonlocal interactions, the ground-state entanglement entropy is known to follow a volume law in the regime where the subsystem size lies below the nonlocality scale \cite{NS_TT, JLKCR, UKCNDSJSMW, DWP}. In the context of holographic entanglement entropy in NCYM theory, the inherent nonlocality is encoded in the deformation of the Ryu--Takayanagi surface \cite{LBCF, WFAKSK, JLKCR, TJZX}. Given this background, it is natural to expect that HSC, together with the associated HFS, may exhibit nontrivial modifications in the presence of nonlocal interactions. In this work, we investigate these effects within the framework of NCYM theory.

Several approaches have been proposed to investigate complexity in noncommutative Yang--Mills theory. For example, Ref.\cite{JCSEWFMLX} analyzed complexity based on the CA conjecture, while Ref. \cite{GKMBSPSRR} addressed complexity in nonlocal field theories and analyzed the subregion complexity of NCYM theory perturbatively in the noncommutativity parameter  and numerically in the non-perturbative regime. In this work, we adopt a complementary approach by numerically studying the HSC and the associated HFS from multiple perspectives.

The paper is organized as follows. In Section 2, we present the HSC conjectured in Ref.\cite{MA}, construct its noncommutative deformation, and derive the universal (cutoff-independent) part for rectangular subregions. We further investigate the scale dependence of this quantity. In Section 3, we review the HFS proposed in Refs.\cite{MMTNNSTTKW, MAAFA}, formulate its noncommutative deformation for rectangular subregions, and analyze its dependence on the noncommutativity parameter. In Section 4, we numerically test the strong subadditivity of HSC in NCYM theory and investigate how its properties differ from those in the commutative theory. In Section 5, we investigate the effects of finite temperature on the HSC and the associated HFS by considering a noncommutatively deformed AdS black hole background. We find that temperature induces various quantitative modifications in the behavior of these quantities. In Section 6, we study these quantities in a noncommutatively deformed AdS soliton background. We observe characteristic signals corresponding to transitions between connected and disconnected configurations in the subregion complexity. Furthermore, we show that there exists a parameter region where the distinction between the commutative and noncommutative theories is significantly reduced. Section 7 is devoted to concluding remarks.

%
%

\section{Holographic subregion complexity in noncommutative Yang-Mills theory} 
\label{sec2}
\setcounter{equation}{0}
\addtocounter{enumi}{1}

We consider a Yang--Mills theory on a noncommutative space, whose subspace is the noncommutative (Moyal) plane defined by a Moyal algebra $[x_{2},\;x_{3}]=i\theta$. Here, $\theta$ is a constant referred to as the noncommutativity parameter, which has the dimension of length squared. The realization of the noncommutative Yang--Mills  theory (NCYM) on D3-branes in a constant NS--NS $B$-field background has a holographic dual in the large $N$, strong ’t Hooft coupling limit. Consider the case where there is one component $B_{23}$ of the NS $B$-field.  This takes the following form \cite{HI, MR, AOSJ, MLYSW} 
\begin{alignat}{2}
\label{201}
ds^{2} =R^{2}
\Bigl[u^{2} \bigl\{& -dx_{0}{}^{2}+dx_{1}{}^{2}  + h(u)(dx_{2}{}^{2}
 + && dx_{3}{}^{2}) \bigr\} 
+ \left(\dfrac{du^{2}}{u^{2}}+d\Omega_{5}{}^{2} \right) \Bigr] \,, \\[2mm]
& B_{23}=R^{2}a^{2}u^{4}h(u)\,, && e^{2\phi}=g_{s}^{2}h(u)\,, 
\nonumber 
\end{alignat}
where $d\Omega_{5}{}^{2}$ denotes the metric on the $S^{5}$ with unit radius. The dilaton field is denoted by $\phi$. The parameter $R$ is given by $R^{4}=4\pi g_{s} N l_{s}^{4}$, where $g_{s}$  is the asymptotic value of the coupling constant, $l_{s}$ is the string length scale, and $N$ is the number of D3-branes. The background Eq.~(\ref{201}) is also characterized by a function, denoted by $h(u)$, which is explicitly given by
\begin{align}
\label{202}
h(u)=\dfrac{1}{1+a^{4}u^{4}} \,,
\end{align}
where the parameter $a$ in Eq.~(\ref{202}), which is the renormalized noncommutativity parameter, has the dimension of length.

The entanglement entropy of the $(d+1)$-dimensional subregion $A$ in the boundary theory is proportional to the area of the so-called Ryu--Takayanagi surface denoted by \cite{SRTT1, SRTT2, LAJM}
\begin{align}
\label{203}
{\cal A}_{\gamma}=\int_{\gamma} d^{d}\sigma \sqrt{G_{\rm int}^{(d)}} \,,
\end{align}
where $\gamma$ is the $d$-dimensional extremal surface in $AdS_{d+2}$ such that the boundary of  $\gamma$ coincides with the boundary of $A$, and $G_{\rm int}^{(d)}$ is the induced string frame metric on $\gamma$. In holographic duals of nonconformal theories like NCYM theory, the dilaton and the volume of the $(8-d)$ compact dimensions are in general not constant. A natural generalization of Eq.~(\ref{203}) is given by \cite{SRTT1, TNTT}
\begin{align}
\label{204}
{\cal A}_{\gamma} &= \int d^{8}\sigma \,e^{-2\phi}\sqrt{G_{\rm int}^{(8)}}\,,
\end{align}
where $G_{\rm int}^{(8)}$ is the determinant of the induced string frame metric in eight dimensions.

We consider the subregion $A$ to be a three-dimensional infinite strip of width $l$ in one direction and width $L (\to \infty)$ in the other two directions. Let us choose these directions as
\begin{align}
\label{205}
x^{2} \in \left[-\dfrac{l}{2}, \,\dfrac{l}{2}\right] \,, \quad 
x^{1}\,,\;x^{3} \in \left[-\dfrac{L}{2}, \,\dfrac{L}{2}\right] \,.
\end{align}
The Ryu--Takayanagi surface for the infinite strip is given by
\begin{align}
\label{206}
{\cal A}_{\gamma} 
= \dfrac{2\pi^{3}L^{2}R^{8}}{g_{s}^{2}}\int du\,u^{3}\sqrt{X'^{2}(u)+\dfrac{1}{u^{4}h(u)}}\,, 
\end{align}
where we have utilized the volume of the unit-radius $S^{5}$, $\Omega_{5}=\pi^{3}$, and redefined the coordinate $x_{2}$ as $X (u)$. The minimal surface condition from Eq.~(\ref{206}) yields the following expression for $X(u)$:
\begin{align}
\label{207}
X(u) &= \int^{u}_{u_{\ast}} dU\,X'(U) \nonumber \\
&= \int^{u}_{u_{\ast}} dU\,\dfrac{1}{U^{2}}
\sqrt{\dfrac{1+a^{4}U^{4}} {\dfrac{U^{6}}{u_{\ast}^{6}}-1}}\;,
\end{align}
where $u_{\ast}$ denotes a constant of motion, and $u=u_{\ast}$ marks the point of closest approach for the extremal surface. The width $l$ is expressed as  $\dfrac{l}{2}=X(u \to \infty)$ and is a function of the parameter $u_{\ast}$. This width $l$ can be regarded as the characteristic length of the subregion. We find that the characteristic length $l$ in NCYM theory has a minimum value of $l_{\rm min} \simeq 1.6\,a$ at $u_{\ast} \simeq 0.79/a$. In other words, unlike in commutative Yang--Mills theory, the inverse function $u_{\ast}=u_{\ast}(l)$ is double-valued for $l>l_{\rm min}$.

For a subregion $A$ in the boundary theory, the holographic subregion complexity is defined as follows \cite{MA, OBADC}\,:
\begin{align}
\label{208}
C_{A} = \dfrac{{\cal V}_{\gamma}}{8\pi G_{N}^{(d+2)}R}\;,
\end{align}
where $R$ is the radius of curvature of $AdS_{d+2}$, $G_{N}^{(d+2)}$ is the (d+2)-dimensional Newton constant, and ${\cal V}_{\gamma}$ is the codimension-one volume of the portion in the bulk geometry enclosed by the minimal Ryu--Takayanagi surface. For the holographic dual of NCYM theory, we consider the natural generalization of Eq.~ (\ref{208}) as: 
\begin{align}
\label{209}
C_{A} = \dfrac{{\cal V}_{\gamma}}{8\pi G_{N}^{(10)}R}\;,
\end{align}
where 
\begin{align}
\label{210}
{\cal V}_{\gamma} &= \int d^{9}\sigma \,e^{-2\phi}
\sqrt{G^{(9)}} \;.
\end{align}
Here $G_{N}^{(10)}=8\pi^{6}l_{s}^{8}$ is the ten-dimensional Newton constant (in the Einstein frame), and $G^{(9)}$ is the determinant of the induced string frame metric in nine dimensions. 
For the three-dimensional infinite strip defined in Eq.~ (\ref{205}), the volume ${\cal V}_{\gamma}$ is given by
\begin{align}
\label{211}
{\cal V}_{\gamma} =\dfrac{2\pi^{3}L^{2}R^{9}}{g_{s}^{2}} \int^{u_{\Lambda}}_{u_{\ast}}du\,u^{2} X(u)\;,
\end{align}
where $u_{\Lambda}$ denotes the ultraviolet cutoff parameter. The volume ${\cal V}_{\gamma}$ diverges as $u_{\Lambda}$ approaches infinity and can be decomposed into two parts: the most divergent part ${\cal V}_{\gamma}^{({\rm div1})}$ and the proper part ${\cal V}_{\gamma}^{({\rm proper})}$\,:
\begin{align}
\label{212}
{\cal V}_{\gamma} 
& = {\cal V}_{\gamma}^{({\rm proper})}+{\cal V}_{\gamma}^{({\rm div1})} \;,
\end{align}
where 
\begin{subequations}
\begin{align}
\label{213a}
{\cal V}_{\gamma}^{({\rm proper})} &
= -\dfrac{2\pi^{3}R^{9}L^{2}}{3g_{s}^{2}} \int^{u_{\Lambda}}_{u_{\ast}} du\,u \sqrt{\dfrac{1+a^{4}u^{4}}{\dfrac{u^{6}}{u_{\ast}^{6}}-1}}\,, \\
& \nonumber  \\[-5mm] 
\label{213b}
{\cal V}_{\gamma}^{({\rm div1})} &= \dfrac{2\pi^{3}R^{9}L^{2}u_{\Lambda}^{3}}{3g_{s}^{2}} 
\int^{u_{\Lambda}}_{u_{\ast}} du\,\dfrac{1}{u^{2}} 
\sqrt{\dfrac{1+a^{4}u^{4}}{\dfrac{u^{6}}{u_{\ast}^{6}}-1}}\,.
\end{align}
\end{subequations}
While the volume ${\cal V}_{\gamma}^{({\rm proper})}$ with a finite noncommutativity parameter remains 
ultraviolet divergent at large $u_{\Lambda}$, it converges in the commutative limit ($a \to 0$). The finite term, independent of the ultraviolet cutoff parameter, can be extracted from ${\cal V}_{\gamma}^{({\rm proper})}$ through the use of recursion formulas.

The finite part of the volume ${\cal V}_{\gamma}^{({\rm proper})}$ can be expressed as a function of the characteristic length $l$. In the following, we examine the behavior of the volume with respect to $l$ in various regimes. 

\begin{description}

\item[Case 1.]  Commutative limit: $a \to 0$.

The volume ${\cal V}_{\gamma}^{({\rm proper})}$ becomes finite as $u_{\Lambda} \to \infty$, and in this limit, it is given by:
\begin{align}
\label{214}
{\cal V}_{\gamma\,{\rm C}}^{\rm (proper)} 
=-\dfrac{\pi^{7/2}R^{9}L^{2}}{9g_{s}^{2}}\dfrac{\Gamma(1/6)}{\Gamma(2/3)}u_{\ast}^{2} \,,
\end{align}
where $\Gamma(x)$ denotes the gamma function. The volume ${\cal V}_{\gamma\,{\rm C}}^{\rm (proper)}$ can be expressed as a function of the characteristic length $l_{\rm C}=2\sqrt{\pi}\,\Gamma(2/3)/u_{\ast}\Gamma(1/6)$. The finite part of the holographic subregion complexity, 
$C_{A\,{\rm C}}^{\rm (univ)} \equiv {\cal V}_{\gamma\,{\rm C}}^{\rm (proper)}/(8\pi G_{N}^{(10)}R)$, 
is independent of the cutoff parameter $u_{\Lambda}$. This quantity can thus be regarded as universal, as it is independent of the regularization scheme. The dependence of $C_{A\,{\rm C}}^{\rm (univ)}$ on the characteristic length $l_{\rm C}$ is given by
\begin{align}
\label{215}
C_{A\,{\rm C}}^{({\rm univ})}
=-\dfrac{N^{2}L^{2}}{9\sqrt{\pi}}
\dfrac{\Gamma(2/3)}{\Gamma(1/6)}\dfrac{1}{\,l_{\rm C}^{2}\,} \,.
\end{align}
Conversely, the dependence of the divergent part, $C_{A\,{\rm C}}^{({\rm div 1})} \equiv {\cal V}_{\gamma\,{\rm C}}^{({\rm div1})}/(8\pi G_{N}^{(10)}R)$, on the characteristic length $l_{\rm C}$ is given by:
\begin{align}
\label{216}
C_{A\,{\rm C}}^{({\rm div 1})} 
=\dfrac{N^{2}}{12\pi^{2}}u_{\Lambda}^{3}(L^{2}l_{\rm C}) \,.
\end{align}
We find that $C_{A\,{\rm C}}^{({\rm div 1})}$ is proportional to the volume of the subregion, $L^{2}l_{\rm C}$.

\item[Case 2.]  Noncommutative limit: $a \gg \dfrac{1}{u_{\ast}}$

Unlike the commutative limit, the volume ${\cal V}_{\gamma}^{({\rm proper})}$ diverges as the cutoff parameter $u_{\Lambda}$ approaches infinity. However,  ${\cal V}_{\gamma}^{({\rm proper})}$ can be expressed as the sum of a finite term, ${\cal V}_{\gamma}^{({\rm finite})}$, and a divergent term, ${\cal V}_{\gamma}^{({\rm div2})}$:
\begin{align}
\label{217}
{\cal V}_{\gamma}^{({\rm proper})} 
= {\cal V}_{\gamma}^{({\rm finite})}+{\cal V}_{\gamma}^{({\rm div2})} \;,
\end{align}
where the terms are
\begin{subequations}
\begin{align}
\label{218a, 218b}
{\cal V}_{\gamma}^{({\rm finite})} &= \dfrac{2\pi^{3} R^{9}L^{2}}{3g_{s}^{2}}\,a^{2}u_{\ast}^{4}\,I \,, \\[3mm]
{\cal V}_{\gamma}^{({\rm div2})} &= -\dfrac{2\pi^{3} R^{9}L^{2}}{3g_{s}^{2}}\,a^{2}u_{\ast}^{3}u_{\Lambda}\, \,.
\end{align}
\end{subequations}
Here, $I$ is defined as
\begin{align}
\label{219}
I=1-\int^{1}_{0}\dfrac{dx}{x^{2}}\left(\dfrac{1}{\sqrt{1-x^{6}}}-1\right)
= 1+\sqrt{\pi}\,\dfrac{\Gamma(5/6)}{\Gamma(1/3)}\,.
\end{align}
The universal term of the holographic subregion complexity, $C_{A}^{({\rm univ})} \equiv {\cal V}_{\gamma}^{({\rm finite})}/(8\pi G_{N}^{(10)}R)$, can be expressed as a function of the characteristic length $l=\sqrt{\pi}\Gamma(1/3)a^{2}u_{\ast}/3\Gamma(5/6)$. Its dependence on $l$ is given by: 
\begin{align}
\label{220}
C_{A}^{({\rm univ})}=\dfrac{27N^{2}}{2\pi^{4}} \dfrac{L^{2}}{a^{2}}
\dfrac{\Gamma(5/6)^{4}}{\Gamma(1/3)^{4}}\,I\, 
\left(\dfrac{l}{a} \right)^{4}\,.
\end{align}
Similarly, the divergent term 
$C_{A}^{({\rm div 1})} \equiv {\cal V}_{\gamma}^{({\rm div1})}/(8\pi G_{N}^{(10)}R)$
can be expressed as:
\begin{align}
\label{221}
C_{A}^{({\rm div 1})} 
=\dfrac{N^{2}}{12\pi^{2}}u_{\Lambda}^{3}(L^{2}l) \,.
\end{align}
We find that $C_{A}^{({\rm div 1})}$ in the noncommutative limit is also proportional to the volume $L^{2}l$.

\end{description}

Let us consider the noncommutative case where the noncommutativity parameter $a$ is in the intermediate regime, $au_{\ast} \sim 1$. Although the volume ${\cal V}_{\gamma}^{({\rm proper})}$ diverges as the cutoff parameter $u_{\Lambda}$ approaches infinity, a finite term ${\cal V}_{\gamma}^{({\rm finite})}$ can be extracted by utilizing the following recursion relations for the binomial integral $I[m,\,n,\,p]=\int x^{m}(cx^{n}+d)^{p}$:
\begin{align}
\label{222}
I[m,\,n,\,p]&=\dfrac{x^{m+1}(cx^{n}+d)^{p}}{m+1}-\dfrac{npc}{m+1}I[m+n,\,n,\,p-1] \nonumber \\
&=\dfrac{x^{m+1}(cx^{n}+d)^{p+1}}{d(m+1)}-\dfrac{c(m+1+n(p+1))}{d(m+1)}I[m+n,\,n,\,p] \;.
\end{align}
Here, $c$ and $d$ are real numbers, and $m\,(\neq -1),\;n$ and $p$ are rational numbers.

Consequently, the volume ${\cal V}_{\gamma}^{({\rm proper})}$ can be expressed as the sum of a finite term ${\cal V}_{\gamma}^{({\rm finite})}$ and a divergent term ${\cal V}_{\gamma}^{({\rm div2})}$:
\begin{align}
\label{223}
{\cal V}_{\gamma}^{({\rm proper})} 
= {\cal V}_{\gamma}^{({\rm finite})}+{\cal V}_{\gamma}^{({\rm div2})} \;,
\end{align}
where these terms are given by:
\begin{subequations}
\begin{align}
\label{224a}
{\cal V}_{\gamma}^{({\rm finite})} 
&= \dfrac{2\pi^{3} R^{9}L^{2}}{3g_{s}^{2}}\,u_{\ast}^{2}\,I_{a}^{(1)} \,, \\[3mm]
{\cal V}_{\gamma}^{({\rm div2})}
\label{224b} 
&= -\dfrac{2\pi^{3} R^{9}L^{2}}{3g_{s}^{2}}\,a^{2}u_{\ast}^{3}u_{\Lambda}\sqrt{1
+\left(\dfrac{1}{au_{\Lambda}}\right)^{4}}\, \,.
\end{align}
\end{subequations}
Here, $I_{a}^{(1)}$ is a function of the dimensionless variable $au_{\ast}$, defined as:
\begin{align}
\label{225}
I_{a}^{(1)} &=\sqrt{1+(au_{\ast})^{4}} 
- 2\int^{1}_{0} dx \dfrac{x^{2}}{\sqrt{x^{4}+(au_{\ast})^{4}}} \\[2mm] \nonumber 
&-\int^{1}_{0} \dfrac{dx}{x^{2}} \, \sqrt{x^{4}+(au_{\ast})^{4}} \, \left(\dfrac{1}{\sqrt{1-x^{6}}}-1\right) \,. 
\end{align}
Alternative expressions for ${\cal V}_{\gamma}^{({\rm finite})}$ and ${\cal V}_{\gamma}^{({\rm div2})}$ are given by:
\begin{subequations}
\begin{align}
\label{226a} 
{\cal V}_{\gamma}^{({\rm finite})} 
&= \dfrac{2\pi^{3} R^{9}L^{2}}{3g_{s}^{2}}\,u_{\ast}^{2}\,I_{a}^{(2)} \,, \\[3mm]
\label{226b}
{\cal V}_{\gamma}^{({\rm div2})} &= -\dfrac{2\pi^{3} R^{9}L^{2}}{3g_{s}^{2}}\,a^{2}u_{\ast}^{3}u_{\Lambda}
\left\{1+\left(\dfrac{1}{au_{\Lambda}}\right)^{4}\right\}^{3/2} \,.
\end{align}
\end{subequations}
Here, $I_{a}^{(2)}$ is defined by:
\begin{align}
\label{227}
I_{a}^{(2)} &=\dfrac{\left\{1+(au_{\ast})^{4}\right\}^{3/2}}{(au_{\ast})^{4}}
- \dfrac{5}{(au_{\ast})^{4}} \int^{1}_{0} dx\, x^{2}\sqrt{x^{4}+(au_{\ast})^{4}} \\[2mm] 
&-\int^{1}_{0} \dfrac{dx}{x^{2}} \, \sqrt{x^{4}+(au_{\ast})^{4}} \, \left(\dfrac{1}{\sqrt{1-x^{6}}}-1\right) 
\,. \nonumber 
\end{align}
The two quantities $I_{a}^{(1)}$ and $I_{a}^{(2)}$ exhibit identical behavior as functions of $au_{\ast}$ and share the same limiting values:
\begin{subequations}
\begin{alignat}{2} 
\label{228a,228b}
\lim_{a \to 0}I_{a}^{(1)} &=\lim_{a \to 0}I_{a}^{(2)}
&&=-\dfrac{\sqrt{\pi}}{6}\dfrac{\Gamma(1/6)}{\Gamma(2/3)} \,, \\[2mm]
\lim_{a \to \infty}I_{a}^{(1)} &=\lim_{a \to \infty}I_{a}^{(2)} &&= (au_{\ast})^{2}\,I \,.
\end{alignat}
\end{subequations}
These limiting values are fully consistent with the results obtained in the commutative and noncommutative regimes, respectively.

\begin{figure}[t]
\centering
\vspace*{0mm}
\includegraphics[width=120mm]{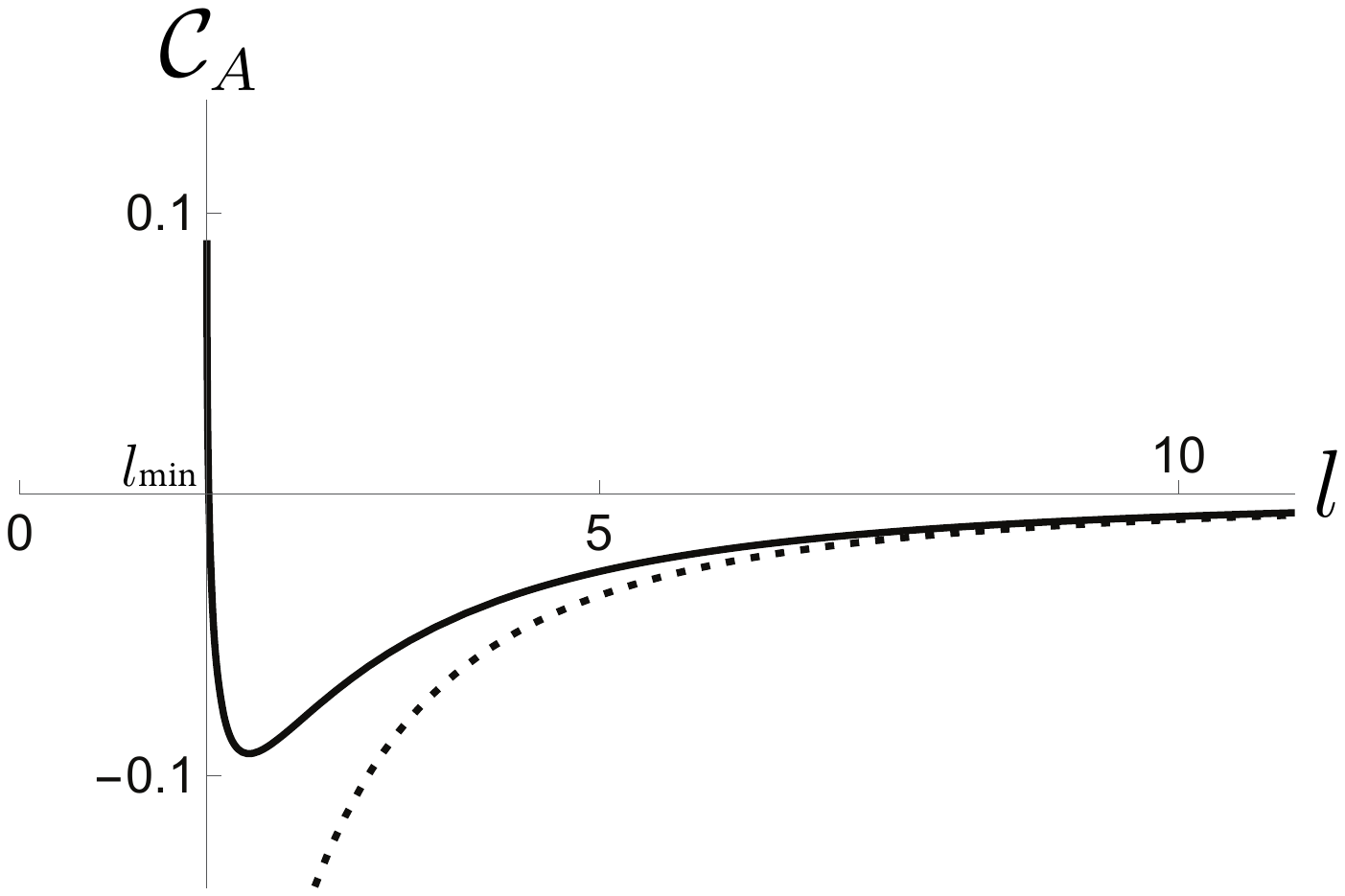} 
\vspace*{-10mm}
\caption{The variation of the dimensionless quantity ${\cal C}_{A} \equiv \dfrac{6\pi^{2}a^{2}}{N^{2}L^{2}}C_{A}^{({\rm univ})}$ with respect to the dimensionless length $l/a$, in units where $a=1$. The solid and dotted lines correspond to the noncommutative case and the commutative limit ($a \to 0$), respectively.}
\label{F1}
\end{figure}

\begin{figure}[t]
\centering
\vspace*{0mm}
\includegraphics[width=120mm]{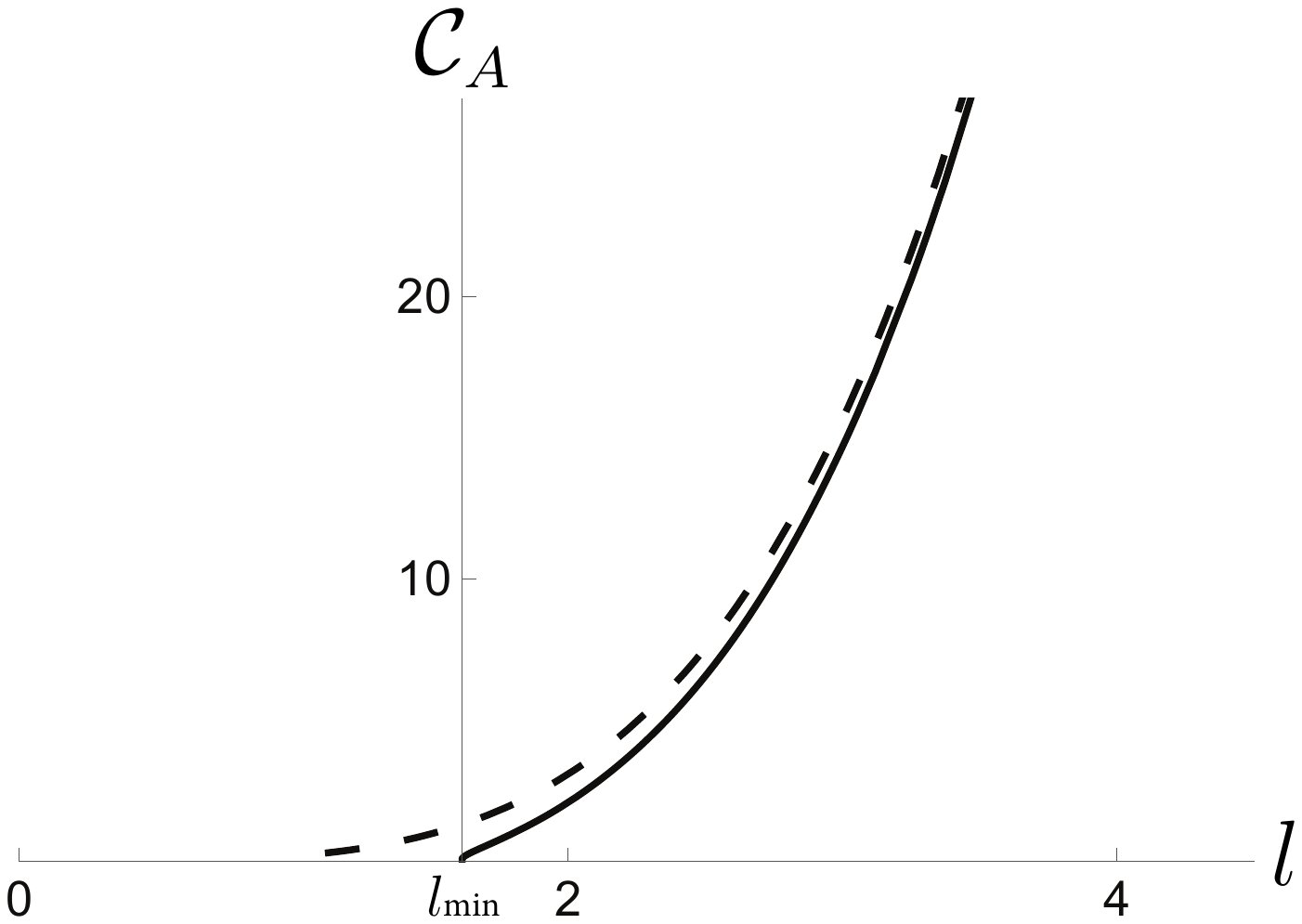} 
\vspace*{-10mm}
\caption{The variation of the dimensionless quantity ${\cal C}_{A} \equiv \dfrac{6\pi^{2}a^{2}}{N^{2}L^{2}}C_{A}^{({\rm univ})}$ with respect to the dimensionless length $l/a$, in units where $a=1$. The solid and dashed lines correspond to the noncommutative case and the noncommutative limit, respectively.}
\label{F2}
\end{figure}

The universal term of the holographic subregion complexity, 
$C_{A}^{({\rm univ})} \equiv {\cal V}_{\gamma}^{({\rm finite})}/(8\pi G_{N}^{(10)}R)$, can also be expressed as a function of the characteristic length $l$. Its numerical dependence on $l$ is illustrated in Fig.\ref{F1} for $au_{\ast} \lesssim 0.79$, and Fig.\ref{F2} for $au_{\ast} \gtrsim 0.79$. Note that the behavior of the universal term $C_{A}^{({\rm univ})}$ as a function of $l$ differs dramatically between the cases $au_{\ast} \lesssim 0.79$ and $au_{\ast} \gtrsim 0.79$.

When the characteristic length $l$ significantly exceeds the minimal length $l_{\rm min}$, the universal term $C_{A}^{({\rm univ})}$ shows only a slight difference between the noncommutative case and the commutative limit. However, this distinction becomes more pronounced as $l$ approaches $l_{\rm min}$. 
Furthermore, at short range scales, the behavior of $C_{A}^{({\rm univ})}$ differs significantly between the noncommutative and commutative regimes. It is particularly noteworthy that the universal term $C_{A}^{({\rm univ})}$ possesses a lower bound in the noncommutative case. 

%
%
\section{Holographic fidelity susceptibility} 
\label{sec3}
\setcounter{equation}{0}
\addtocounter{enumi}{1}

In this section, we discuss the relationship between holographic subregion complexity and fidelity susceptibility, and investigate the effects of noncommutative deformations on the latter. Consider a one-parameter family of quantum (pure) states denoted by $\ket{\psi(\rho)}$, where $\rho$ is a parameter.  When we infinitesimally perturb the parameter from $\rho$ to $\rho + \delta \rho$, the state vector becomes $\ket{\psi(\rho + \delta \rho)}$. The fidelity susceptibility, denoted by $G_{\rho}$, is then defined by the inner product of the two states as: 
\begin{align}
\label{301}
\left| \braket{\psi(\rho)}{\psi(\rho + \delta \rho)} \right| 
= 1-G_{\rho} \cdot (\delta \rho)^{2} + O((\delta \rho)^{3})\,. 
\end{align}
The quantity $G_{\rho}$ is also called the (quantum) information metric or Bures metric. It measures the distance between two infinitesimally different quantum states \cite{SJG}.

A holographic estimation of the fidelity susceptibility for quantum states of a $(d+1)$-dimensional boundary CFT was proposed in \cite{MMTNNSTTKW}. According to this proposal, the fidelity susceptibility can be estimated by the maximal volume of a codimension-one bulk surface:
\begin{align}
\label{302}
G_{\rho} =\dfrac{{\rm Vol}(\Sigma_{max})}{R^{d+1}}\,,
\end{align}
up to an $O(1)$ constant. Here, $\Sigma_{max}$ is a $(d+1)$-dimensional spacelike bulk surface that ends on a time slice at the boundary of the AdS space and has the maximal volume in the AdS space.

\begin{figure}[t]
\centering
\vspace*{0mm}
\includegraphics[width=120mm]{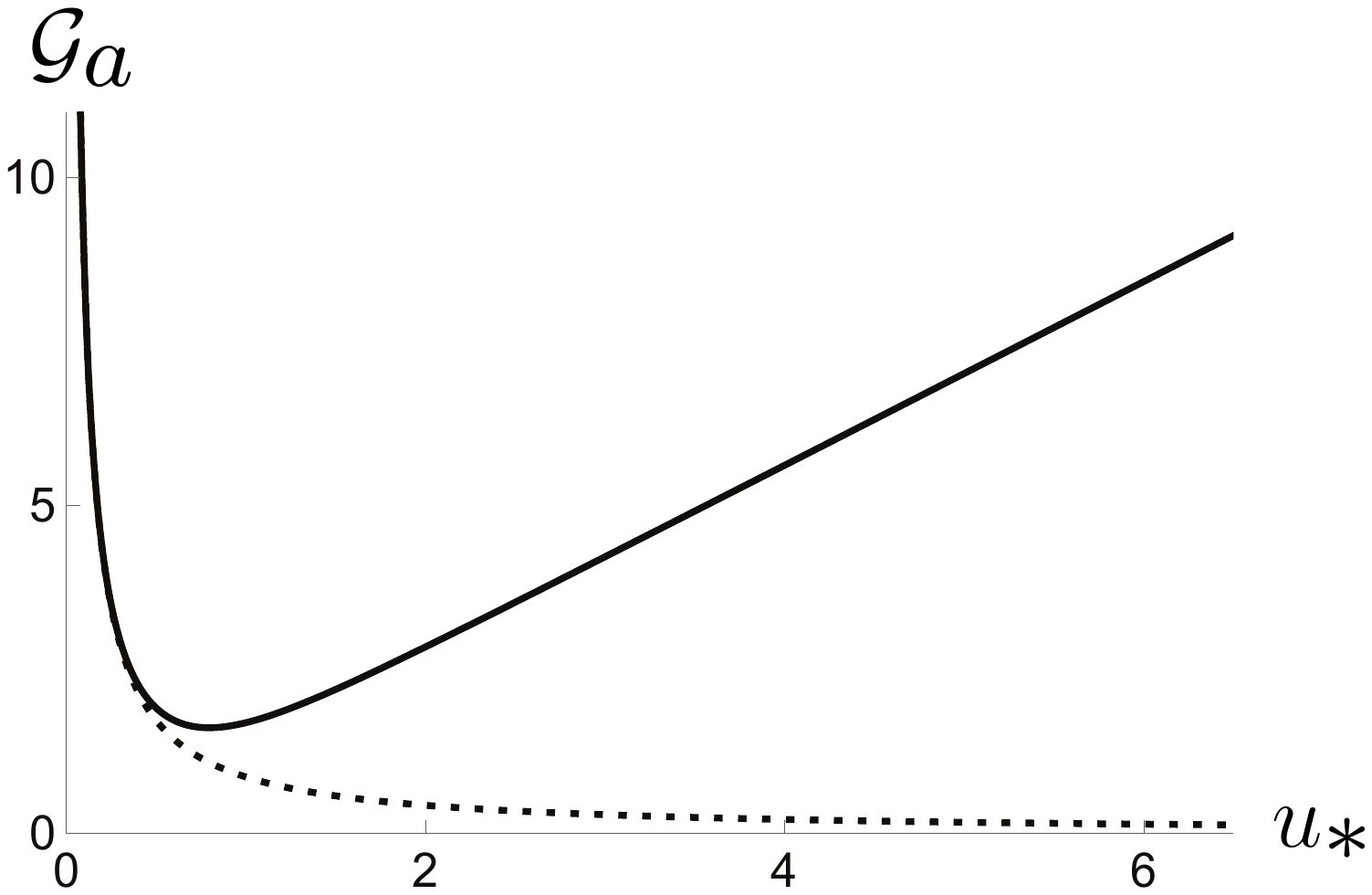} 
\vspace*{-10mm}
\caption{The variation of the dimensionless quantity ${\cal G}_{a} \equiv \dfrac{12\pi^{2}}{au_{\Lambda}^{3}N^{2}L^{2}} G_{a}$ as a function of the parameter $au_{\ast}$, in units where $a=1$. 
The solid and dotted lines correspond to the noncommutative case and the commutative limit, respectively.}
\label{F3}
\end{figure}

From the perspective of the holographic subregion complexity, its most divergent part is presumed to correspond to the holographic fidelity susceptibility $G_{\rho}$ in the dual description of the field theory \cite{MA}. By interpreting $C_{A}^{({\rm div 1})} = {\cal V}_{\gamma}^{({\rm div1})}/(8\pi G_{N}^{(10)}R)$ 
as the holographic fidelity susceptibility $G_{a}$ in the holographic dual description of the noncommutative Yang--Mills theory, we obtain:  
\begin{align}
\label{303}
G_{a} = \dfrac{N^{2}}{12\pi^{2}}u_{\Lambda}^{3}L^{2}l \,,
\end{align}
where the characteristic length is given by $l=2X(u\!\to\!\infty)$. The numerical evaluation of the dependence of $G_{a}$ on the parameter $u_{\ast}$ is shown in Fig.\ref{F3}. 
Interpreting $u_{\ast}$ as an inverse scale parameter, we find that the holographic fidelity susceptibility behaves very differently in the noncommutative theory compared to the commutative theory at short-range scales.
Fig.\ref{F3} reveals that, unlike the commutative case, the holographic fidelity susceptibility $G_{a}$ in the noncommutative Yang--Mills theory possesses a non-zero lower bound.

\begin{figure}[t]
\centering
\vspace*{0mm}
\includegraphics[width=120mm]{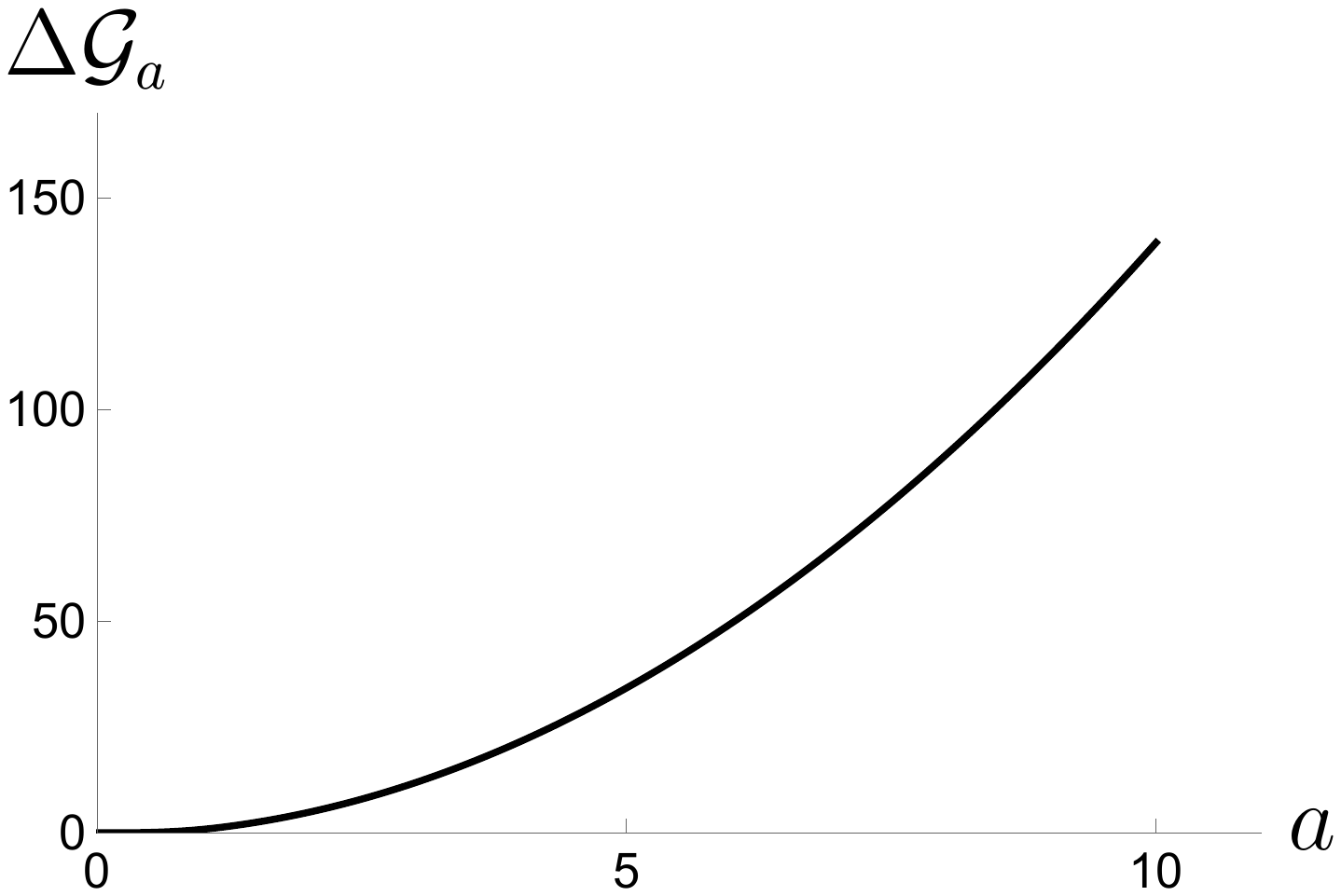} 
\vspace*{-10mm}
\caption{The variation of the dimensionless quantity $\Delta{\cal G}_{a} \equiv \dfrac{12\pi^{2}u_{\ast}}{u_{\Lambda}^{3}N^{2}L^{2}} \Delta G_{a}$ as a function of the dimensionless parameter $au_{\ast}$, in units where $u_{\ast}=1$.}
\label{F4}
\end{figure}

The noncommutative Yang--Mills theory can be regarded as a one-parameter deformation of the ordinary Yang--Mills theory, characterized by the noncommutativity parameter. From this perspective, we define the regularized holographic fidelity susceptibility of the noncommutative Yang--Mills theory, denoted by $\Delta G_{a}$, as the difference between its holographic fidelity susceptibility and that in the commutative limit,  obtained by taking the noncommutativity parameter to zero \cite{MAAFA, KBDMMAA}:
\begin{align}
\label{304}
\Delta G_{a} = \dfrac{N^{2}}{12\pi^{2}}u_{\Lambda}^{3}L^{2}\Delta l\,,
\end{align}
where $\Delta l \equiv l-l_{\rm C}$. The numerical evaluation of the regularized holographic fidelity susceptibility $\Delta G_{a}$ as a function of the dimensionless parameter $au_{\ast}$ is illustrated in Fig.\ref{F4}. 
As shown in Fig.\ref{F4}, the regularized holographic fidelity susceptibility, $\Delta G_{a}$, of the noncommutative Yang--Mills theory increases monotonically as a function of the noncommutativity parameter $a$.

%
%
\section{Strong subadditivity}

\setcounter{equation}{0}
\addtocounter{enumi}{1}

There are several important properties that entanglement entropy generally satisfies. One such property is strong subadditivity (SSA), expressed by the following inequality:
\begin{align}
\label{401}
S_{A}+S_{B}-S_{A \cup B}-S_{A \cap B} \geq 0 \,,
\end{align}
where the subscripts $A$ and $B$ denote any two subsystems,  and $S_{A}$ and $S_{B}$ represent their respective entanglement entropies. It has been shown that holographic entanglement entropy, which is proportional to the area of the Ryu--Takayanagi surface, also satisfies the SSA inequality \cite{MHTT, TNSRTT}.

\begin{figure}[t]
\centering
\vspace*{-20mm}
\hspace*{0mm}
\includegraphics[width=150mm]{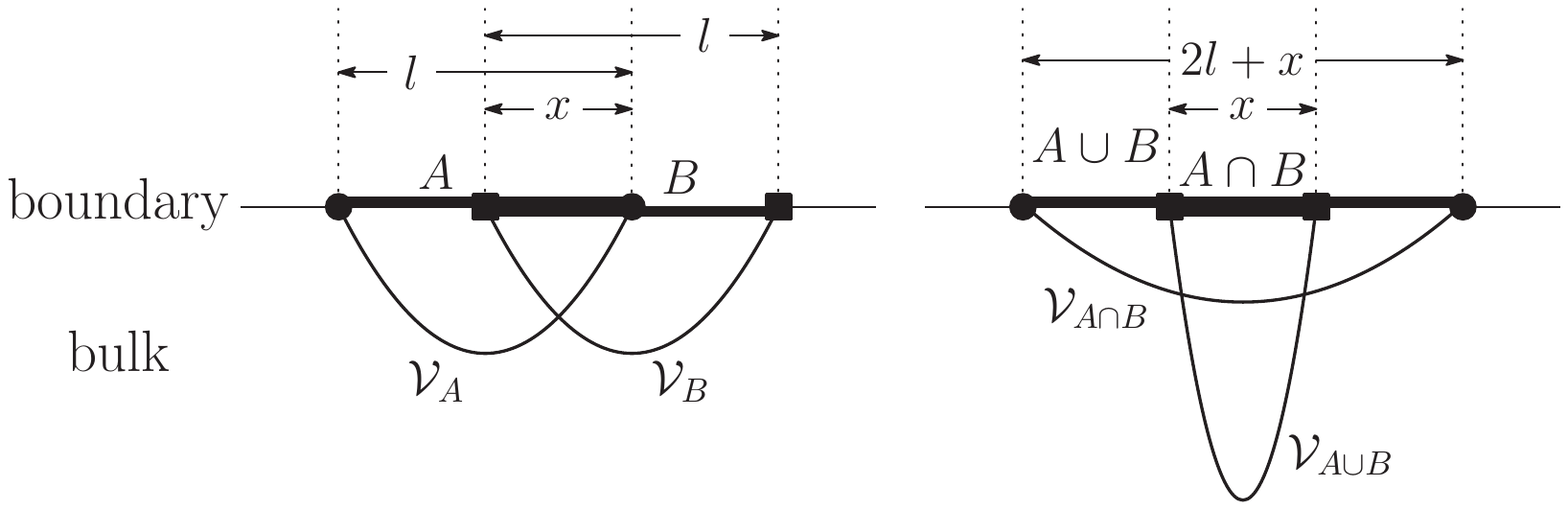}
\vspace*{-25mm}
\caption{(Left) Two overlapping infinite boundary strips $A$ and $B$, with their respective volumes ${\cal V}_{A}$ and ${\cal V}_{B}$ enclosed by the Ryu--Takayanagi surface. (Right) Two overlapping infinite boundary strips $A \cup B$ and $A \cap B$, with their respective volumes ${\cal V}_{A \cup B}$ and ${\cal V}_{A \cap B}$ enclosed by the Ryu--Takayanagi surface.}
\label{F5}
\end{figure}

Naively, one might expect that the holographic subregion complexity, which is proportional to the volume enclosed by the Ryu--Takayanagi surface, would also satisfy strong subadditivity. Eq.~(\ref{215}) shows that the holographic subregion complexity in ordinary Yang--Mills theory (the commutative limit of the noncommutative theory) is an inverse square function of $l_{\rm C}$, where $l_{\rm C}$ denotes the characteristic length of the subregion. It is straightforward to verify from the form of Eq.~(\ref{215}) that the holographic subregion complexity in ordinary Yang--Mills theory satisfies SSA. However, it is not immediately obvious whether this property holds in noncommutative Yang--Mills theory. In what follows, we numerically investigate the validity of SSA in NCYM for a specific case. Consider two infinite strips (subregions) $A$ and $B$, each with a characteristic length $l$, overlapping with a width $x$ along the $x_{2}$-direction. A schematic illustration is given in Fig.\ref{F5}.

\begin{figure}[t]
\centering
\vspace*{0mm}
\includegraphics[width=120mm]{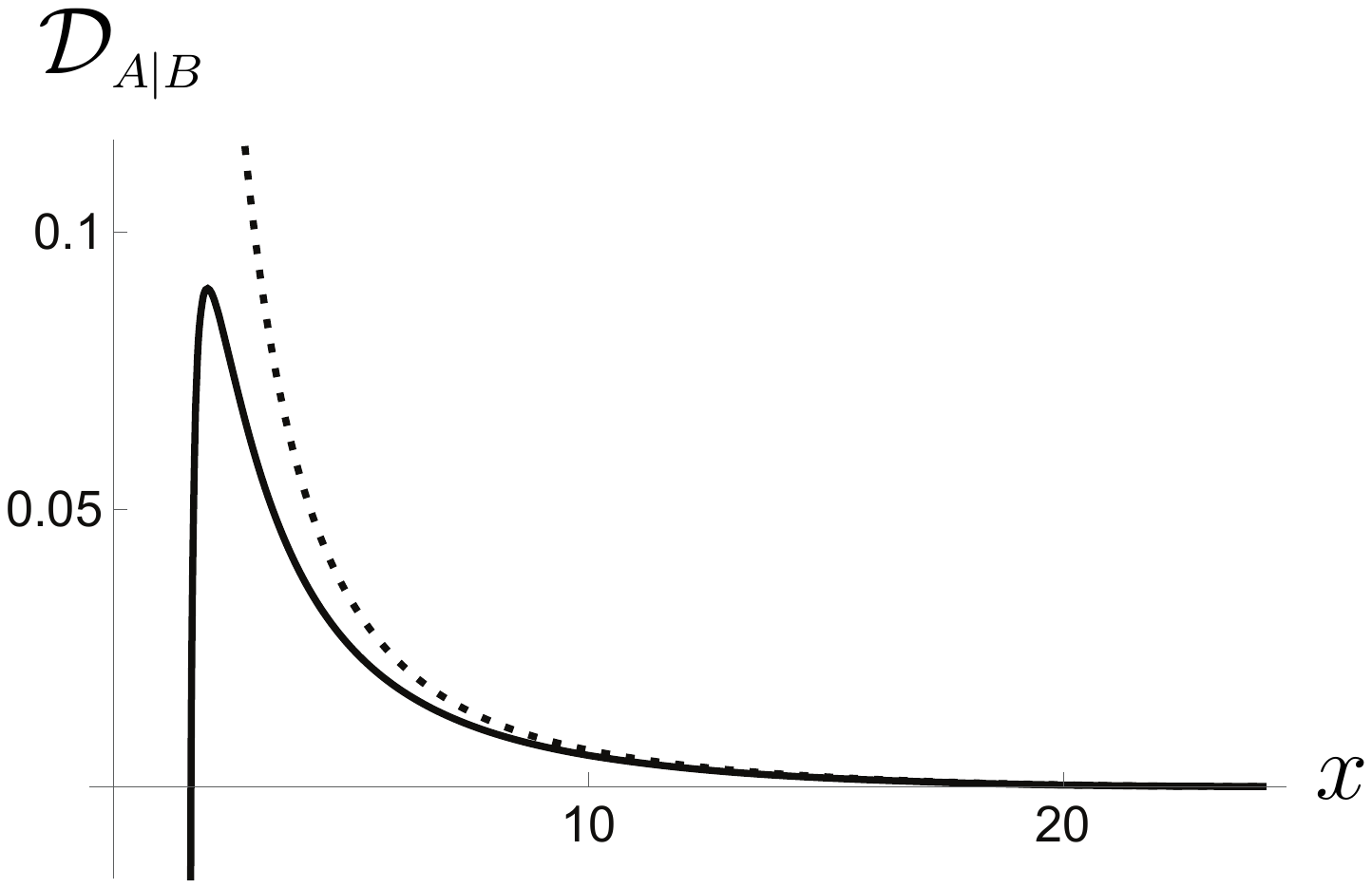} 
\vspace*{-10mm}
\caption{The behavior of the dimensionless quantity ${\cal D}_{A|B} \equiv \dfrac{6\pi^{2}a^{2}}{N^{2}L^{2}}D_{A|B}$ as a function of the width $ax$, in units where $a=1$. The range of the width $x$ is $l_{\rm min} \leq x \leq 15\,l_{\rm min}$. The solid line corresponds to the noncommutative case ($a=1$), while the dotted line corresponds to the commutative limit ($a \to 0$).}
\label{F6}
\end{figure}

Let us define the following quantity:
\begin{align}
\label{402}
D_{A|B} \equiv C_{A}^{({\rm univ})}+C_{B}^{({\rm univ})}
-C_{A \cup B}^{({\rm univ})}-C_{A \cap B}^{({\rm univ})} \,.
\end{align}
This quantity can be regarded as a function of the width $x$ of the overlapping intervals. As discussed in Sec.\ref{sec2}, the characteristic length $l$ is bounded from below by $l_{\rm min}$. Consequently, the width $x$ is constrained to the interval $l_{\rm min} \leq x \leq l$. If the quantity $D_{A|B}$ is positive in the interval $l_{\rm min} \leq x \leq l$, then the holographic subregion complexity in noncommutative Yang--Mills theory is said to satisfy strong subadditivity. The numerical evaluation of the dependence of  $D_{A|B}$ on $x$ is shown in Fig.\ref{F6}. 
Fig. \ref{F6} reveals that while there is little difference in behavior between the noncommutative case and the commutative limit in the large-$x$ region ($x \gg l_{\rm min}$), a striking discrepancy emerges as the width $x$ approaches $l_{\rm min}$. While the quantity $D_{A|B}$ remains positive for most of the range, ensuring the validity of strong subadditivity, it exhibits a sharp plunge and becomes negative in the immediate vicinity of $l_{\rm min}$. This indicates that the holographic subregion complexity in noncommutative Yang--Mills theory satisfies strong subadditivity except in the very close proximity of the minimum length $l_{\rm min}$. The existence of this lower bound $l_{\rm min}$ fundamentally alters the short-distance entanglement structure, leading to a localized violation of the naive SSA-like property for complexity. 

%
%

\section{Finite temperature} 

\setcounter{equation}{0}
\addtocounter{enumi}{1}

In this section, we examine the holographic subregion complexity in noncommutative Yang–-Mills theory at finite temperature. To this end, we analyze the noncommutative deformation of the AdS black hole background. The corresponding metric is expressed as a modified form of Eq.~(\ref{201}):
\begin{align}
\label{501}
ds^{2} &=R^{2}
\Bigl[u^{2} \bigl\{f_{T}(u)\,dx_{E}{}^{2}+dx_{1}{}^{2}  + h(u)(dx_{2}{}^{2}
 + dx_{3}{}^{2}) \bigr\} 
+ \left(\dfrac{du^{2}}{u^{2}f_{T}(u)}+d\Omega_{5}{}^{2} \right) \Bigr] \,, 
\end{align}
where the blackening factor is given by $f_{T}=1-\left(\dfrac{u_{T}}{u} \right)^{4}$and $u_{T}$ is a parameter with dimension of mass. In deriving the background metric (\ref{501}), we introduce the Euclidean time $x_{E}=ix_{0}$. The temperature $T$ of the noncommutative Yang--Mills theory is proportional to the parameter $u_{T}$, as given by $T = u_{T}/\pi$. The volume given by Eq.~(\ref{211}) is then modified as
\begin{align}
\label{502}
{\cal V}_{\gamma T} =\dfrac{2\pi^{3}L^{2}R^{9}}{g_{s}^{2}} \int^{u_{\Lambda}}_{u_{\ast}}du\,u^{2} 
X_{T}(u)\;, 
\end{align}
where $X_{T}(u)$ is defined as 
\begin{align}
\label{503}
X_{T}(u) 
& =\int^{u}_{u_{\ast}} dU\,\dfrac{1}{U^{2}\sqrt{1-\left(\dfrac{u_{T}}{U}\right)^{4}}}
\sqrt{\dfrac{1+a^{4}U^{4}} {\dfrac{U^{6}}{u_{\ast}^{6}}-1}}\;,
\end{align}
The characteristic length $l_{T}$ is defined by $\dfrac{l_{T}}{2}=X_{T}(u \to \infty)$. As evident from its definition, $l_{T}$ is a function of the parameters $u_{\ast}$ and $u_{T}$, and it coincides with the characteristic length $l$ in the limit $u_{T} \to 0$.

We set the magnitude of the noncommutativity parameter to approximately $au_{\ast} \sim 1$. The volume ${\cal V}_{\gamma T}$ can be decomposed into a finite term ${\cal V}_{\gamma T}^{(\rm finite)}$ and two divergent terms, ${\cal V}_{\gamma T}^{(\rm div 1)}$ and ${\cal V}_{\gamma}^{(\rm div 2)}$, as follows:
\begin{align}
\label{504}
{\cal V}_{\gamma T} 
& = {\cal V}_{\gamma T}^{(\rm finite)}
+{\cal V}_{\gamma T}^{(\rm div1)} +{\cal V}_{\gamma}^{(\rm div2)} \;.
\end{align}
Here, the divergent term ${\cal V}_{\gamma}^{(\rm div2)}$ is given by Eq.~(\ref{224b}), while the term ${\cal V}_{\gamma}^{(\rm div1)}$ is expressed as
\begin{align}
\label{505}
{\cal V}_{\gamma T}^{({\rm div1})} &= \dfrac{2\pi^{3}R^{9}L^{2}u_{\Lambda}^{3}}{3g_{s}^{2}} 
\int^{u_{\Lambda}}_{u_{\ast}} du\,\dfrac{1}{u^{2}\sqrt{1-\left(\dfrac{u_{T}}{u}\right)^{4}}}
\sqrt{\dfrac{1+a^{4}u^{4}}{\dfrac{u^{6}}{u_{\ast}^{6}}-1}}\,.
\end{align}
The finite term ${\cal V}_{\gamma T}^{(\rm finite)}$ can be written as
\begin{align}
\label{506}
{\cal V}_{\gamma T}^{(\rm finite)} = \dfrac{2\pi^{3} R^{9}L^{2}}{3g_{s}^{2}}\,a^{2}u_{\ast}^{4} 
\,(I_{\alpha T}^{(1)} + J_{\alpha T}) \,.
\end{align}
Here, $I_{\alpha T}^{(1)}$ and $J_{\alpha T}$ are functions of the dimensionless quantities $au_{\ast}$ and $u_{T}/u_{\ast}$, defined as follows: 
\begin{subequations}
\begin{align}
\label{507a}
I_{\alpha T}^{(1)} &=\sqrt{1+(au_{\ast})^{4}} 
- 2\int^{1}_{0} dx \dfrac{x^{2}}{\sqrt{x^{4}+(au_{\ast})^{4}}} \\[2mm] \nonumber 
&-\int^{1}_{0} \dfrac{dx}{x^{2}} \, \sqrt{x^{4}+(au_{\ast})^{4}} \, 
\Biggl(\dfrac{1}{\sqrt{(1-x^{6})
\left\{1-\left(\dfrac{u_{T}}{u_{\ast}}x \right)^{4}\right\}}}-1\Biggr) \,. \\
& \nonumber  \\[-5mm]
\label{507b}
J_{\alpha T} &=3\int^{1}_{0} \dfrac{x}{\sqrt{1-\left(\dfrac{u_{T}}{u_{\ast}}x \right)^{4}}} \sqrt{\dfrac{x^{4}+(au_{\ast})^{4}}{1-x^{6}}} 
\int^{x}_{0} \dfrac{dy}{y^{4}} \, \Biggl(\dfrac{1}{\sqrt{
1-\left(\dfrac{u_{T}}{u_{\ast}}x \right)^{4}}}-1\Biggr) \,.
\end{align}
\end{subequations}
It is straightforward to verify that $\lim_{u_{T} \to 0} I_{\alpha T}^{(1)}$ coincides with $I_{\alpha}^{(1)}$, as given by  Eq.~(\ref{225}), and that $\lim_{u_{T} \to 0} J_{\alpha T}$ evaluates to zero. The finite term, ${\cal V}_{\gamma T}^{(\rm finite)}$ can be expressed in alternative forms depending on the definition of the divergent term ${\cal V}_{\gamma}^{(\rm div 2)}$. When ${\cal V}_{\gamma}^{(\rm div 2)}$ is given by Eq.~(\ref{226b}), instead of  Eq.~(\ref{224b}), ${\cal V}_{\gamma T}^{(\rm finite)}$ is given by 
\begin{align}
\label{508}
{\cal V}_{\gamma T}^{(\rm finite)} = \dfrac{2\pi^{3} R^{9}L^{2}}{3g_{s}^{2}}\,a^{2}u_{\ast}^{4} 
\,(I_{\alpha T}^{(2)} + J_{\alpha T}) \,,
\end{align}
where $J_{\alpha T}$ is provided by Eq.~(\ref{507b}) and $I_{\alpha T}^{(2)}$ is defined as follows:
\begin{align}
\label{509}
I_{\alpha T}^{(2)} &=\dfrac{\left\{1+(au_{\ast})^{4}\right\}^{3/2}}{(au_{\ast})^{4}}
- \dfrac{5}{(au_{\ast})^{4}} \int^{1}_{0} dx\, x^{2}\sqrt{x^{4}+(au_{\ast})^{4}} \\[2mm]  
&-\int^{1}_{0} \dfrac{dx}{x^{2}} \, \sqrt{x^{4}+(au_{\ast})^{4}} \, 
\Biggl(\dfrac{1}{\sqrt{(1-x^{6})
\left\{1-\left(\dfrac{u_{T}}{u_{\ast}}x \right)^{4}\right\}}}-1\Biggr) \,. \nonumber
\end{align}
Similar to $\lim_{u_{T} \to 0}I_{\alpha T}^{(1)}=I_{\alpha}^{(1)}$, the limit $\lim_{u_{T} \to 0} I_{\alpha T}^{(2)}$ coincides with $I_{\alpha}^{(2)}$, as given by Eq.~(\ref{227}).

\begin{figure}[t]
\centering
\vspace*{0mm}
\includegraphics[width=120mm]{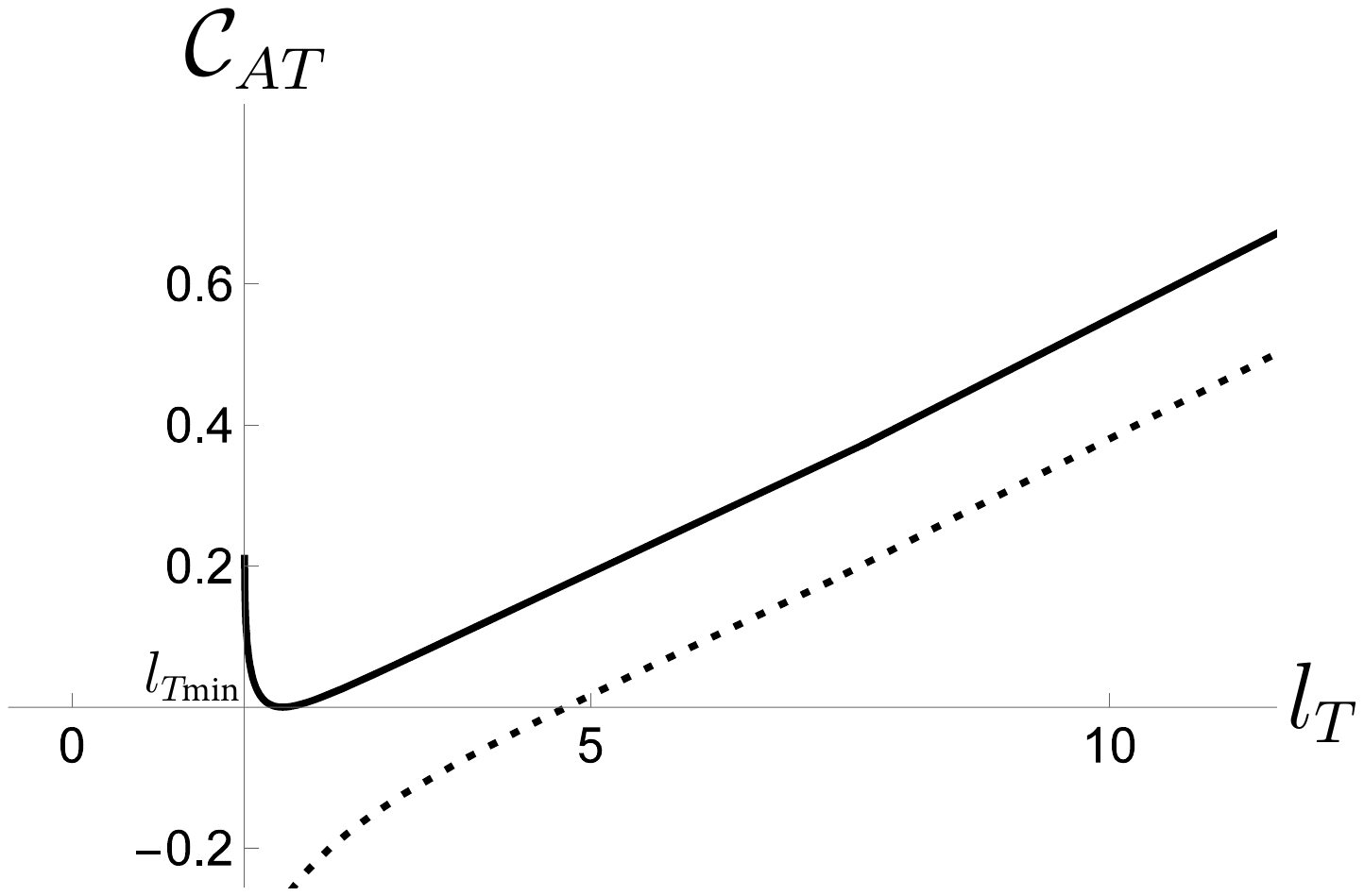} 
\vspace*{-10mm}
\caption{Plot of the dimensionless quantity ${\cal C}_{AT} \equiv \dfrac{6\pi^{2}a^{2}}{N^{2}L^{2}}C_{AT}^{({\rm univ})}$ as a function of the dimensionless length $l_{T}$, evaluated at $au_{T}=0.47245$ in units where $a=1$. 
The solid line corresponds to the noncommutative theory, whereas the dotted line indicates the commutative limit ($a \to 0$). The label $l_{T{\rm min}}$ indicates the minimum value of  $l_{T}$ in the noncommutative theory.}
\label{F7}
\end{figure}

\begin{figure}[t]
\centering
\vspace*{0mm}
\includegraphics[width=120mm]{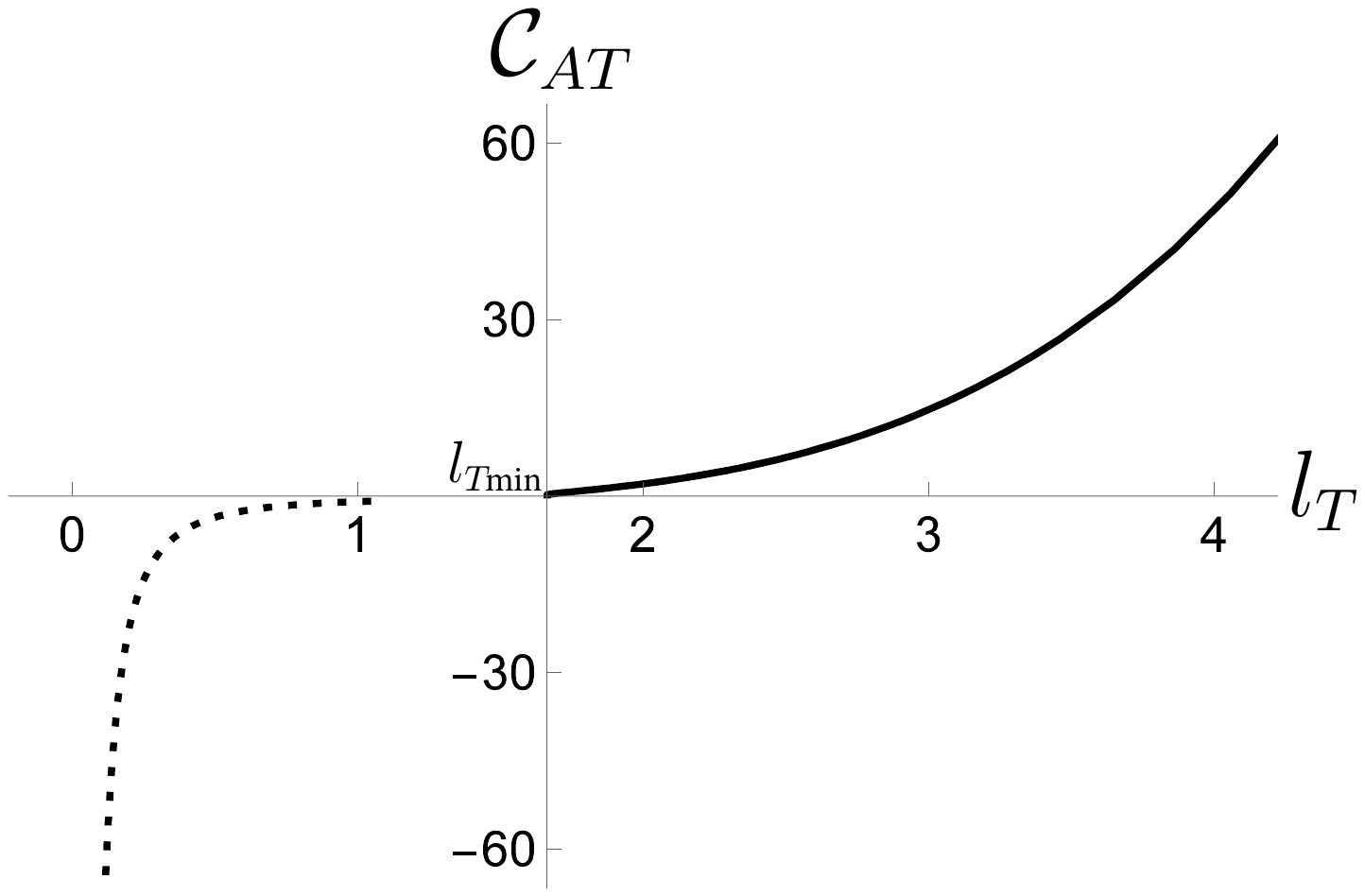} 
\vspace*{-10mm}
\caption{Plot of the dimensionless quantity ${\cal C}_{AT} \equiv \dfrac{6\pi^{2}}{N^{2}L^{2}}C_{AT}^{({\rm univ})}$ as a function of the dimensionless length $l_{T}$, for $au_{T}=0.47245$ in units where $a=1$. 
The solid line corresponds to the noncommutative theory, whereas the dotted line indicates the commutative limit ($a \to 0$).}
\label{F8}
\end{figure}

The universal term of the holographic subregion complexity, 
$C_{AT}^{({\rm univ})} \equiv {\cal V}_{\gamma T}^{({\rm finite})}/(8\pi G_{N}^{(10)}R)$, can also be expressed as a function of the characteristic length $l_{T}$. The numerical dependence of  $C_{AT}^{({\rm univ})}$ on $l_{T}$ is illustrated in Fig.\ref{F7} for $au_{\ast} \lesssim 0.8497$, and in Fig.\ref{F8} for $au_{\ast} \gtrsim 0.8497$. 
As shown in Fig.\ref{F7}, the universal term of the holographic subregion complexity at finite temperature,   $C_{AT}^{({\rm univ})}$, exhibits a lower bound, analogous to the behavior observed in the zero-temperature case. This lower bound increases monotonically with the parameter $u_{T}$, which is associated with the temperature, and becomes zero at $au_{T} \approx 0.47245$, denoted as $au_{T0}$. As discussed in Sec. \ref{sec2}, at zero temperature, the universal term asymptotically approaches zero from negative values as the characteristic length increases. At low temperatures ($u_{T}<u_{T0}$), $C_{AT}^{({\rm univ})}$ transitions from negative to positive as the characteristic length increases. In contrast, at high temperatures ($u_{T}>u_{T0}$), it remains positive across the entire range of the characteristic length. The influence of temperature becomes increasingly significant in the large-length regime. 
As shown in Fig.\ref{F8}, the universal term $C_{AT}^{({\rm univ})}$ in the noncommutative theory 
increases without an upper bound as the characteristic length grows; this behavior is similar to that shown in Fig.\ref{F7}. In contrast,  in the commutative limit, $C_{AT}^{({\rm univ})}$ decreases without a lower bound as the characteristic length approaches zero. This distinction is reminiscent of UV/IR mixing, which is a  characteristic feature of the noncommutative field theories \cite{MRS}.

\begin{figure}[t]
\centering
\vspace*{0mm}
\includegraphics[width=120mm]{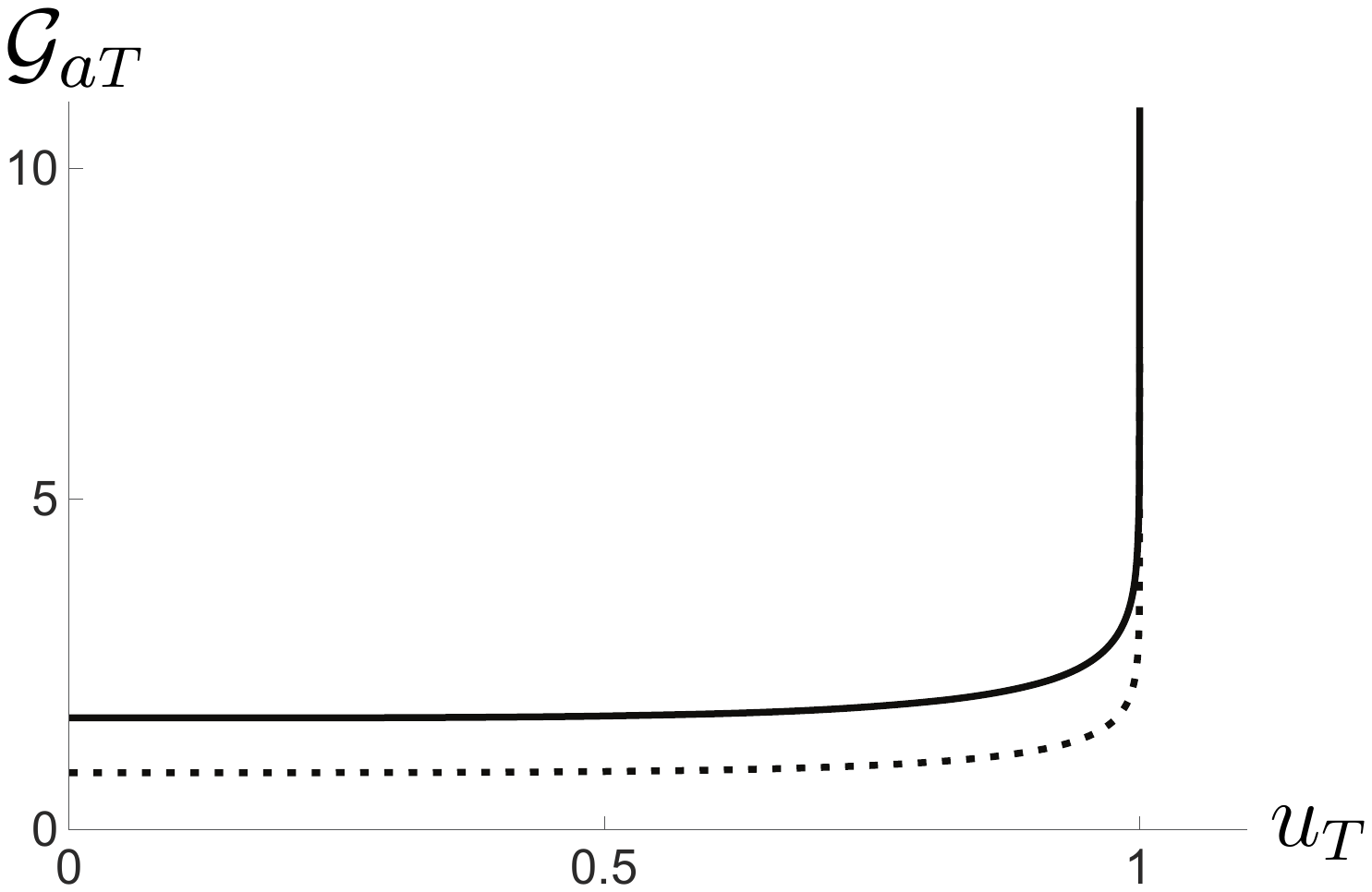} 
\vspace*{-10mm}
\caption{Plot of the dimensionless quantity ${\cal G}_{aT} \equiv \dfrac{12\pi^{2}u_{\ast}}{u_{\Lambda}^{3}N^{2}L^{2}}G_{aT}$ as a function of the dimensionless parameter $u_{T}/u_{\ast}$, in units where $u_{\ast}=1$. The noncommutativity parameter is held fixed at $a=1$. The solid and dotted lines correspond to the noncommutative case, and the commutative limit, respectively.}
\label{F9}
\end{figure}

We now turn our attention to the holographic fidelity susceptibility at finite temperature. Utilizing Eqs.~ (\ref{503}) and (\ref{505}), we derive the explicit expression for the quantity $G_{aT}$, defined as the holographic fidelity susceptibility at finite temperature ${\cal V}_{\gamma T}^{({\rm div1})}/(8\pi G_{N}^{(10)}R)$, as
\begin{align}
\label{510}
G_{aT} = \dfrac{N^{2}}{12\pi^{2}}u_{\Lambda}^{3}L^{2}l_{T}\,.
\end{align}
The numerical evaluation of the dependence of $G_{aT}$ on the parameter $u_{T}/u_{\ast}$ is illustrated in Fig.\ref{F9}. 
At finite temperature, the holographic fidelity susceptibility is largely insensitive to $u_T$ in both the noncommutative theory and its commutative limit, except in the vicinity of $u_T \simeq u_{\ast}$ (with $u_{\ast}=1$). Away from this region, the two results differ in magnitude over the entire range of $u_T$. In both cases, the susceptibility exhibits a sharp peak at $u_T \simeq u_{\ast}$, indicative of phase-transition-like behavior.

\begin{figure}[t]
\centering
\vspace*{0mm}
\includegraphics[width=120mm]{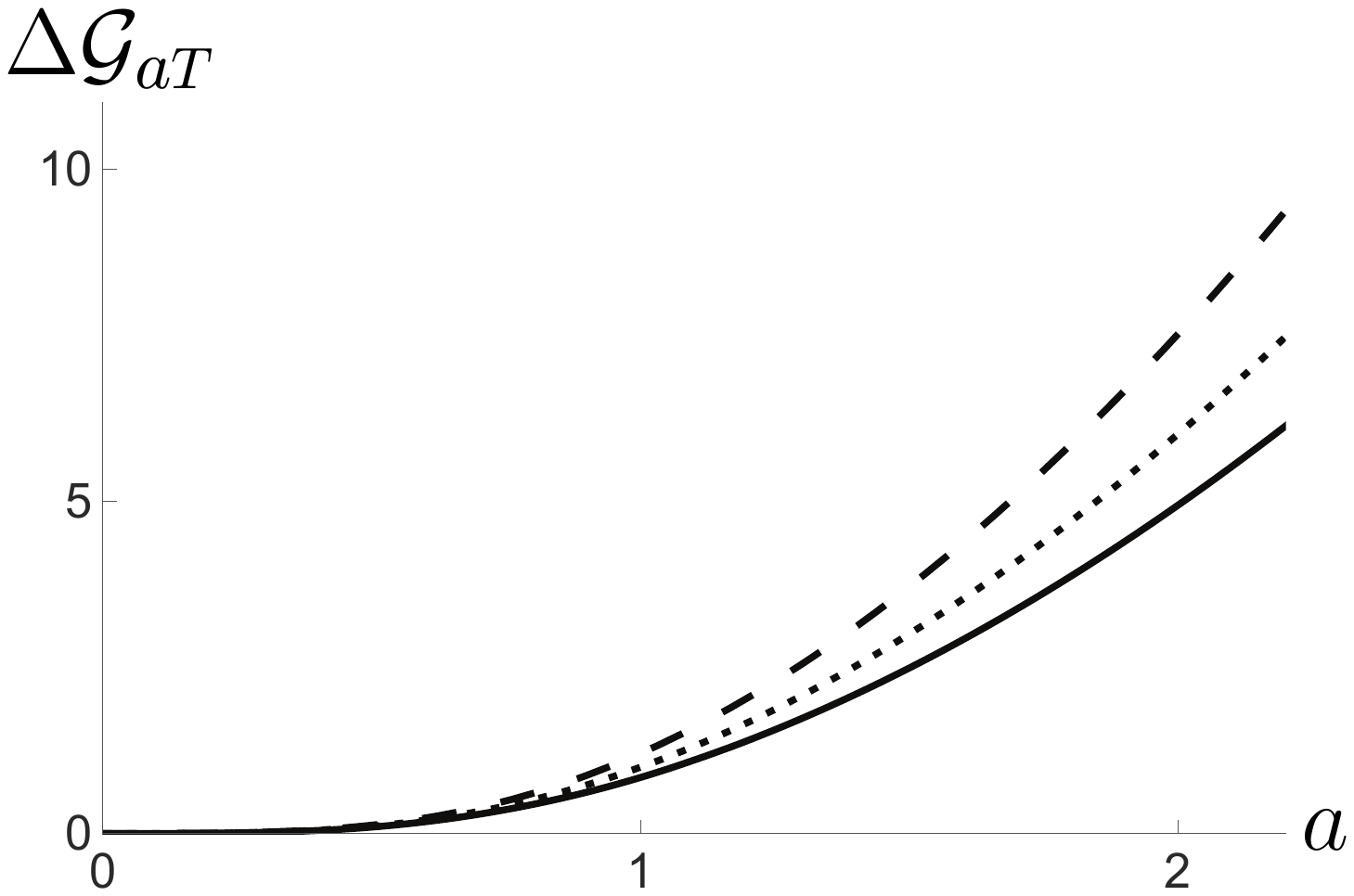} 
\vspace*{-10mm}
\caption{Plot of the dimensionless quantity $\Delta {\cal G}_{aT} \equiv \dfrac{12\pi^{2}u_{\ast}}{u_{\Lambda}^{3}N^{2}L^{2}}\Delta G_{aT}$ as a function of the dimensionless parameter $au_{\ast}$, in units where $u_{\ast}=1$. The solid, dotted, and dashed lines correspond to a low-temperature case ($(u_{T}/u_{\ast})^{4}=0.1$), an intermediate-temperature case ($(u_{T}/u_{\ast})^{4}=0.65$), and a high-temperature case ($(u_{T}/u_{\ast})^{4}=0.9$), respectively.}
\label{F10}
\end{figure}

On the other hand, the regularized holographic fidelity susceptibility at finite temperature for the noncommutative Yang--Mills theory, denoted by $\Delta G_{aT}$, is given by 
\begin{align}
\label{511}
\Delta G_{aT} = \dfrac{N^{2}}{12\pi^{2}}u_{\Lambda}^{3}L^{2}(l_{T}-l_{\rm TC})\,,
\end{align}
where $l_{\rm TC}$ represents the commutative limit of the length $l_{T}$. The numerical dependence of  $\Delta G_{aT}$ on the dimensionless parameter $au_{\ast}$ is illustrated in Fig.\ref{F10}. 
The regularized holographic fidelity susceptibility at finite temperature increases monotonically with the noncommutativity parameter $a$, regardless of the magnitude of the parameter $u_{T}$. Notably, the rate of this increase becomes steeper as $u_{T}$ increases.  In this analysis, the parameter $u_{T}$ is restricted to the range $0 < u_{T}/u_{\ast} < 1$. As shown in Fig.\ref{F10}, this suggests that while temperature affects the quantitative details, it does not alter the overall qualitative behavior of the regularized holographic fidelity susceptibility.

%
%
\section{AdS soliton background} 

\setcounter{equation}{0}
\addtocounter{enumi}{1}

In the previous section, we investigated the noncommutative deformation of the AdS black hole background to examine the holographic subregion complexity in the noncommutative Yang--Mills theory at finite temperature. In this section, we turn our attention to the noncommutative deformation of the AdS soliton  background \cite{GTHRCM}, which is obtained from the double Wick-rotated form of the metric (\ref{501}). The AdS soliton background is widely used to analyze phase transitions within the framework of the gauge/gravity correspondence \cite{IRKDKAM, TSSS1, TSSS2}. This background metric is given by:
\begin{align}
\label{601}
ds^{2} &=R^{2}
\Bigl[u^{2} \bigl\{-dx{}^{2}+f_{K\!K}(u)\,dx_{1}{}^{2}  + h(u)(dx_{2}{}^{2}
 + dx_{3}{}^{2}) \bigr\} 
+ \left(\dfrac{du^{2}}{u^{2}f_{K\!K}(u)}+d\Omega_{5}{}^{2} \right) \Bigr] \,, 
\end{align}
where $f_{K\!K}(u) \equiv 1-\left(\dfrac{u_{K\!K}}{u} \right)^{4}$, with $u_{K\!K}$ being a parameter with the dimension of mass. The radius $r_{p}$ of the compactified circle is related to $u_{K\!K}$ by $r_{p}=1/(2u_{K\!K})$. The volume given in Eq.~(\ref{211}) is modified as 
\begin{align}
\label{602}
{\cal V}_{\gamma K\!K} =\dfrac{2\pi^{4}LR^{9}}{g_{s}^{2}u_{K\!K}} \int^{u_{\Lambda}}_{u_{\ast}}du\,u^{2} 
X_{K\!K}(u)\;, 
\end{align}
where $X_{K\!K}(u)$ is defined as 
\begin{align}
\label{603}
X_{K\!K}(u) 
& =\int^{u}_{u_{\ast}} dU\,\dfrac{1}{U^{2}\sqrt{1-\left(\dfrac{u_{K\!K}}{U}\right)^{4}}}
\sqrt{\dfrac{1+a^{4}U^{4}} {\dfrac{U^{6}\left(1-\left(\dfrac{u_{K\!K}}{U}\right)^{4}\right)}
{u_{\ast}^{6}\left(1-\left(\dfrac{u_{K\!K}}{u_{\ast}}\right)^{4}\right)}-1}}\;.
\end{align}
The characteristic length, $l_{K\!K}$, defined by $l_{K\!K}/2=X_{K\!K}(u \to \infty)$, depends on the parameters $u_{\ast}$ and $u_{K\!K}$. In analogy with the behavior of the characteristic length $l_{T}$, $l_{K\!K}$ reduces to the length $l$ in the limit $u_{K\!K} \to 0$. We also observe that the derivative of $X_{K\!K}(u)$ with respect to $u$ vanishes when the parameters $u_{\ast}$ and $u_{K\!K}$ coincide. This behavior indicates the existence of a disconnected Ryu--Takayanagi surface in addition to the connected configuration.

Following the approach in the preceding section, we now examine the case where the noncommutativity parameter satisfies $au_{\ast} \sim 1$. The volume ${\cal V}_{\gamma K\!K}$ can be decomposed into a finite part ${\cal V}_{\gamma K\!K}^{(\rm finite)}$ and two divergent parts, ${\cal V}_{\gamma K\!K}^{(\rm div 1)}$ and ${\cal V}_{\gamma K\!K}^{(\rm div 2)}$, as described by the following expression:
\begin{align}
\label{604}
{\cal V}_{\gamma K\!K} 
& = {\cal V}_{\gamma K\!K}^{(\rm finite)}
+{\cal V}_{\gamma K\!K}^{(\rm div1)} +{\cal V}_{\gamma K\!K}^{(\rm div2)} \;.
\end{align}
Here, the divergent term ${\cal V}_{\gamma K\!K}^{(\rm div1)}$ is expressed as
\begin{align}
\label{605} 
{\cal V}_{\gamma K\!K}^{({\rm div1})} &= \dfrac{2\pi^{4}R^{9}Lu_{\Lambda}^{3}}{3g_{s}^{2}u_{K\!K}} 
\int^{u_{\Lambda}}_{u_{\ast}} du\,\dfrac{1}{u^{2}\sqrt{1-\left(\dfrac{u_{K\!K}}{u}\right)^{4}}}
\sqrt{\dfrac{1+a^{4}u^{4}}{\dfrac{u^{6}}{u_{\ast}^{6}}-1}}\,. 
\end{align}
The divergent term, ${\cal V}_{\gamma K\!K}^{(\rm div2)}$, admits the following two equivalent expressions:  
\begin{align}
\label{606}
{\cal V}_{\gamma K\!K}^{({\rm div2})} 
= \begin{cases}
-\dfrac{2\pi^{4}R^{9}L}{3g_{s}^{2}u_{K\!K}} \,a^{2}u_{\ast}\,\sqrt{u_{\ast}^{4}-u_{K\!K}^{4}}
\;\sqrt{1+\left(\dfrac{1}{au_{\Lambda}}\right)^{4}} \\[2mm]
-\dfrac{2\pi^{4}R^{9}L}{3g_{s}^{2}u_{K\!K}} \,a^{2}u_{\ast}\,\sqrt{u_{\ast}^{4}-u_{K\!K}^{4}}
\;\left\{1+\left(\dfrac{1}{au_{\Lambda}}\right)^{4}\right\}^{3/2} 
\end{cases} \,.
\end{align}
Corresponding to these two forms of ${\cal V}_{\gamma K\!K}^{(\rm div2)}$, the finite term ${\cal V}_{\gamma K\!K}^{(\rm finite)}$ can also be expressed in two equivalent ways:
\begin{align}
\label{607}
{\cal V}_{\gamma K\!K}^{(\rm finite)} = \begin{cases}
\dfrac{2\pi^{3} R^{9}L}{3g_{s}^{2}u_{K\!K}}\,\sqrt{u_{\ast}^{4}-u_{K\!K}^{4}} \,I_{\alpha K\!K}^{(1)} \\[4mm]
\dfrac{2\pi^{3} R^{9}L}{3g_{s}^{2}u_{K\!K}}\,\sqrt{u_{\ast}^{4}-u_{K\!K}^{4}} \,I_{\alpha K\!K}^{(2)}  
\end{cases} \,.
\end{align}
Here, $I_{\alpha K\!K}^{(1)}$ and $I_{\alpha K\!K}^{(2)}$ are functions of the dimensionless quantities $au_{\ast}$ and $u_{K\!K}/u_{\ast}$, defined as follows: 
\begin{subequations} \label{eq:group}
\begin{align}
I_{\alpha K\!K}^{(1)} &=\sqrt{1+(au_{\ast})^{4}} 
- 2\int^{1}_{0} dx \dfrac{x^{2}}{\sqrt{x^{4}+(au_{\ast})^{4}}} \nonumber \\[2mm]  
&
-\int^{1}_{0} \dfrac{dx}{x^{2}} \, \sqrt{x^{4}+(au_{\ast})^{4}} \label{eq:608a} \\ 
& \qquad \Biggl(\dfrac{1}{\sqrt{\left\{1-\left(\dfrac{u_{K\!K}}{u_{\ast}}x \right)^{4}-
x^{6}\left(1-\left(\dfrac{u_{K\!K}}{u_{\ast}} \right)^{4}\right)\right\}
\left\{1-\left(\dfrac{u_{K\!K}}{u_{\ast}}x \right)^{4}\right\}}}-1\Biggr) \,, \nonumber \\[5mm] 
I_{\alpha K\!K}^{(2)} &=\dfrac{\left\{1+(au_{\ast})^{4}\right\}^{3/2}}{(au_{\ast})^{4}}
- \dfrac{5}{(au_{\ast})^{4}} \int^{1}_{0} dx\, x^{2}\sqrt{x^{4}+(au_{\ast})^{4}} \nonumber \\[2mm]  
&
-\int^{1}_{0} \dfrac{dx}{x^{2}} \, \sqrt{x^{4}+(au_{\ast})^{4}} \label{eq:608b} \\
& \qquad \Biggl(\dfrac{1}{\sqrt{\left\{1-\left(\dfrac{u_{K\!K}}{u_{\ast}}x \right)^{4}-
x^{6}\left(1-\left(\dfrac{u_{K\!K}}{u_{\ast}} \right)^{4}\right)\right\}
\left\{1-\left(\dfrac{u_{K\!K}}{u_{\ast}}x \right)^{4}\right\}}}-1\Biggr) \,, \nonumber 
\end{align}
\end{subequations}
It is also straightforward to verify that, in the limit $u_{K\!K} \to 0$, the functions $I_{\alpha K\!K}^{(1)}$ and  $I_{\alpha K\!K}^{(2)}$ reduce to $I_{\alpha}^{(1)}$ and  $I_{\alpha}^{(2)}$, respectively.

\begin{figure}[t]
\centering
\vspace*{0mm}
\includegraphics[width=120mm]{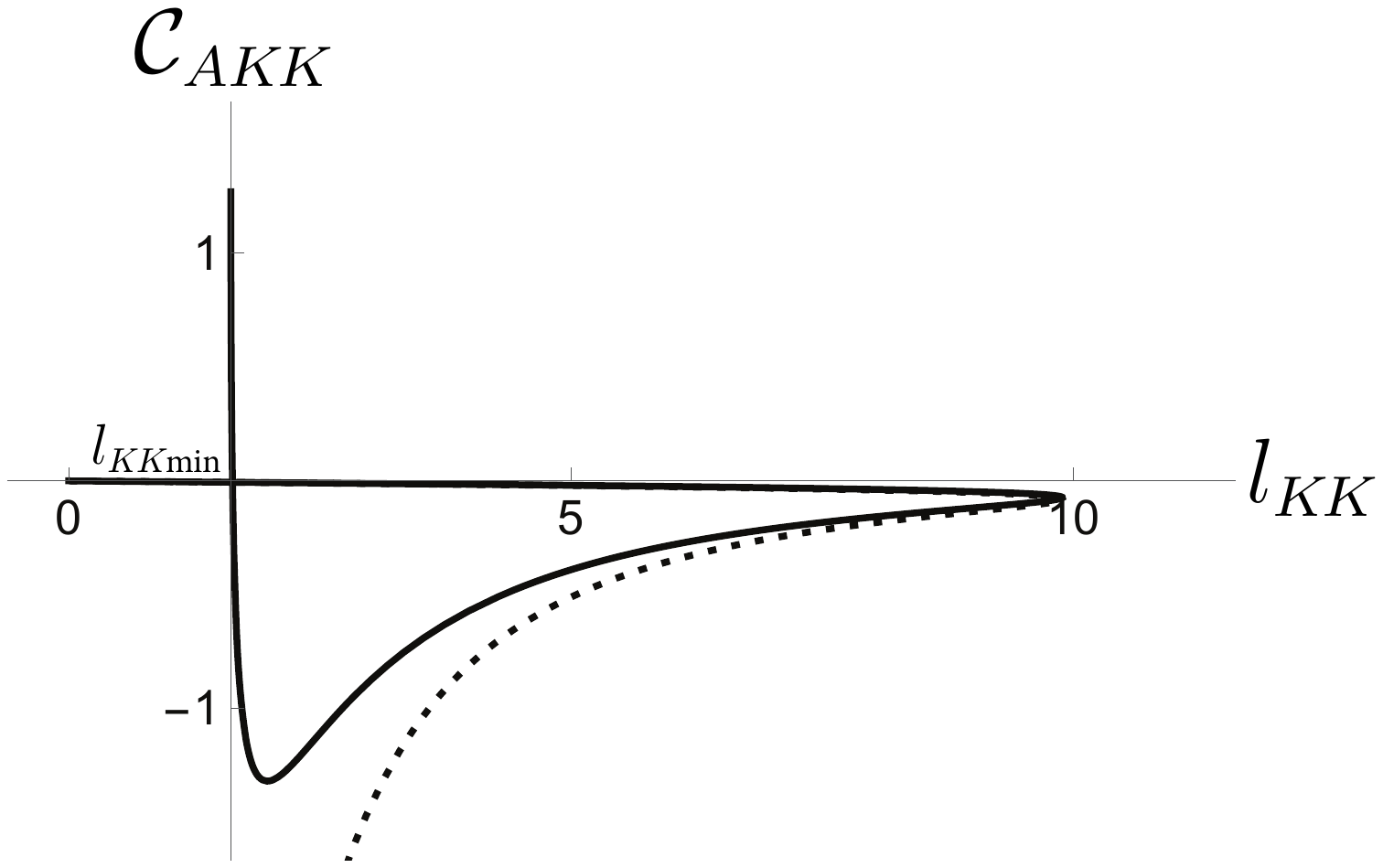} 
\vspace*{-10mm}
\caption{The dimensionless quantity ${\cal C}_{AK\!K} \equiv \dfrac{6\pi^{2}a^{2}}{N^{2}Lu_{K\!K}}C_{AK\!K}^{({\rm univ})}$ versus the dimensionless length $l_{K\!K}/a$, evaluated at $au_{K\!K}=0.07$ in units where $a=1$. The solid line corresponds to the noncommutative theory, whereas the dotted line indicates the commutative limit ($a \to 0$). The label $l_{K\!K{\rm min}}$ indicates the minimum value of  $l_{K\!K}$ in the noncommutative theory.}
\label{F11}
\end{figure}

\begin{figure}[t]
\centering
\vspace*{0mm}
\includegraphics[width=120mm]{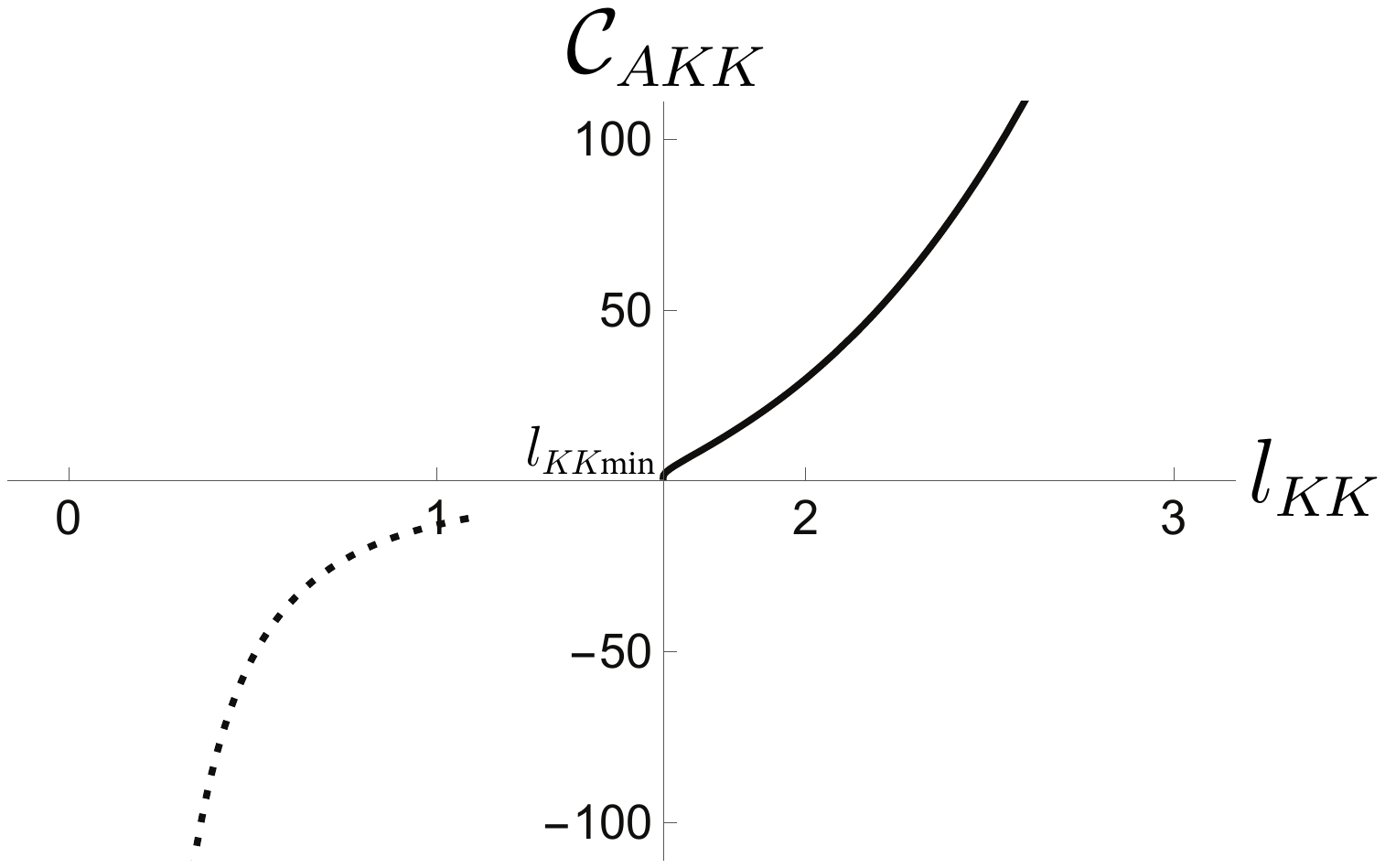} 
\vspace*{-10mm}
\caption{The dimensionless quantity ${\cal C}_{AK\!K} \equiv \dfrac{6\pi^{2}a^{2}}{N^{2}Lu_{K\!K}}C_{AK\!K}^{({\rm univ})}$ versus the dimensionless length $l_{K\!K}/a$, evaluated at $au_{K\!K}=0.07$ in units where $a=1$. The solid line corresponds to the noncommutative theory, whereas the dotted line indicates the commutative limit ($a \to 0$).}
\label{F12}
\end{figure}

The numerical evaluation of the dependence of the universal term of the holographic subregion complexity, $C_{AK\!K}^{({\rm univ})} \equiv {\cal V}_{\gamma K\!K}^{({\rm finite})}/(8\pi G_{N}^{(10)}R)$, on the characteristic length $l_{K\!K}$ is illustrated in Fig.\ref{F11} for $au_{\ast} \lesssim 0.79456$, and in Fig.\ref{F12} for $au_{\ast} \gtrsim 0.79456$. 
%

As noted above, the Ryu--Takayanagi surface admits a disconnected configuration at $u_{\ast}=u_{K\!K}$. 
In this phase, both the characteristic length $l_{K\!K}$ and the universal term of the holographic subregion complexity, $C_{AK\!K}^{({\rm univ})}$, vanish in the disconnected configuration. As shown in Fig.\ref{F11},  $l_{K\!K}$ abruptly drops from its maximal length to zero in the neighborhood of $u_{\ast} = u_{K\!K}$. Accordingly, $l_{K\!K}$ has not only a minimum length $l_{K\!K {\rm min}}$ but also a maximum length in the noncommutative theory. Since the disconnected configuration yields $C_{AK\!K}^{(\rm{univ})}=0$, the value of $C_{AK\!K}^{(\rm{univ})}$ computed on the connected configuration can be interpreted as the difference relative to the disconnected configuration. As shown in Fig.\ref{F11}, $C_{AK\!K}^{(\mathrm{univ})}$ changes sign in the neighborhood of the minimal length $l_{K\!K}^{\min}$. Interpreting $C_{AK\!K}^{(\mathrm{univ})}$ as an order-parameter-like quantity, this sign reversal signals a transition between the connected and disconnected configurations. 

On the other hand, as illustrated in Fig.\ref{F12}, the universal term $C_{AK\!K}^{({\rm univ})}$ grows without bound in the noncommutative theory as the characteristic length increases, whereas it diverges negatively in the commutative limit as the length tends toward zero. The characteristic feature of the noncommutative theory, namely UV/IR mixing, is clearly manifest in this case as well.

\begin{figure}[t]
\centering
\vspace*{0mm}
\includegraphics[width=120mm]{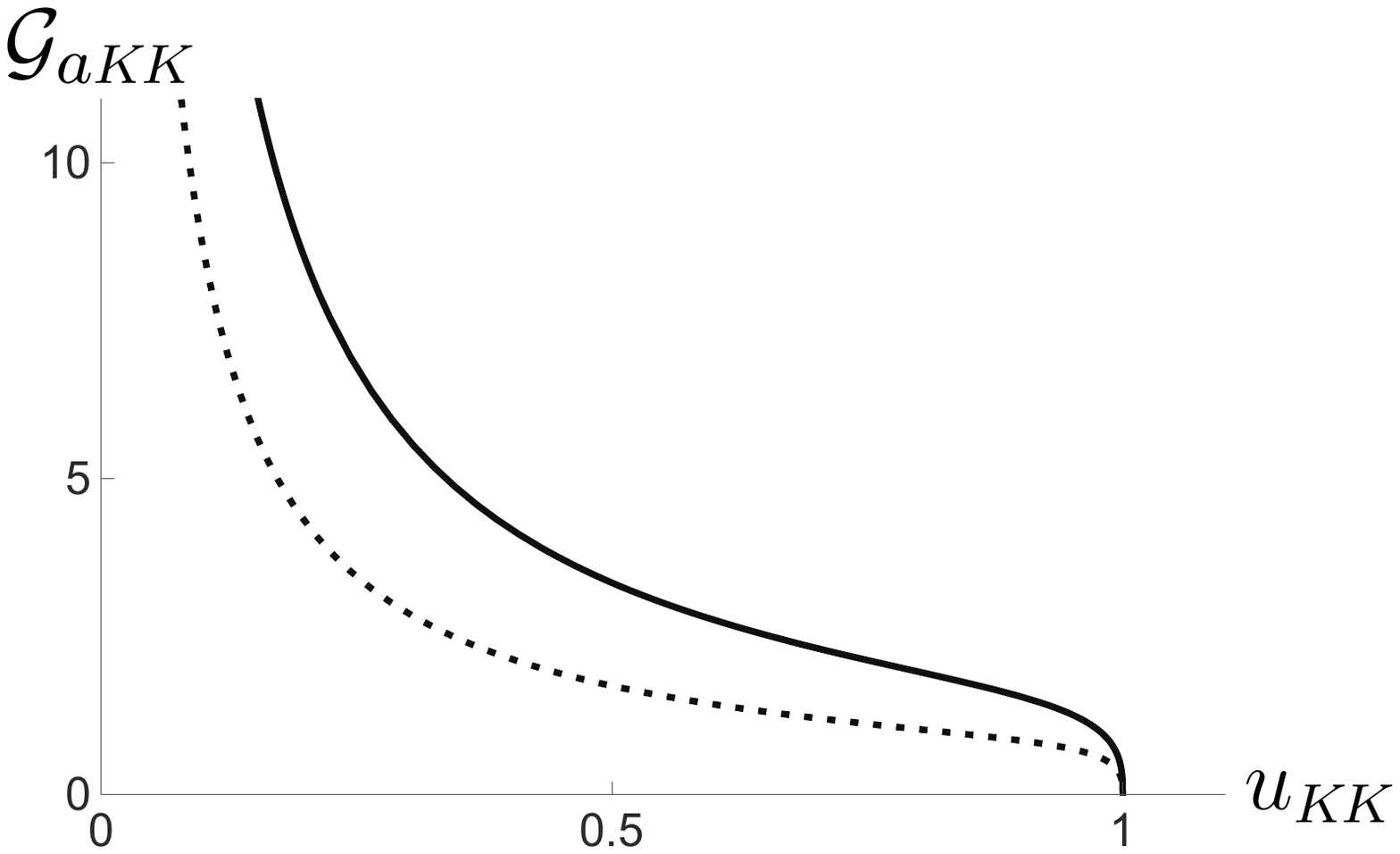} 
\vspace*{-10mm}
\caption{Plot of the dimensionless quantity ${\cal G}_{aK\!K} \equiv \dfrac{12\pi^{2}u_{\ast}^{2}}{u_{\Lambda}^{3}N^{2}L}G_{aK\!K}$ as a function of the dimensionless parameter $u_{K\!K}/u_{\ast}$, in units where $u_{\ast}=1$. The noncommutativity parameter is held fixed at $a=1$. The solid and dotted lines correspond to the noncommutative case and the commutative limit, respectively.}
\label{F13}
\end{figure}

The analysis proceeds to the holographic fidelity susceptibility in the noncommutative deformation of the AdS soliton background. Making use of Eqs.~(\ref{603}) and (\ref{605}), we derive the explicit  expression for the quantity $G_{aK\!K}$, which is defined as ${\cal V}_{\gamma K\!K}^{({\rm div1})}/(8\pi G_{N}^{(10)}R)$, representing the holographic fidelity susceptibility in this background, as 
\begin{align}
\label{609}
G_{aK\!K} = \dfrac{N^{2}}{12\pi^{2}}
\dfrac{u_{\Lambda}^{3}}{u_{K\!K}}Ll_{K\!K}\,.
\end{align}
The numerical evaluation of the dependence of $G_{aK\!K}$ on the parameter $u_{K\!K}/u_{\ast}$ is illustrated in Fig.\ref{F13}. 
In the reduced theory, the holographic fidelity susceptibility is sensitive to variations in the parameter $u_{K\!K}$ and shows the same qualitative dependence in both the noncommutative case and its commutative limit. In either case, the susceptibility develops a gradual peak as $u_{K\!K} \to 0$, corresponding to the transition from the reduced to the unreduced theory. By contrast, the susceptibility vanishes as $u_{K\!K} \to u_{\ast}$, since in the disconnected configuration its value is identically zero.

\begin{figure}[t]
\centering
\vspace*{0mm}
\includegraphics[width=120mm]{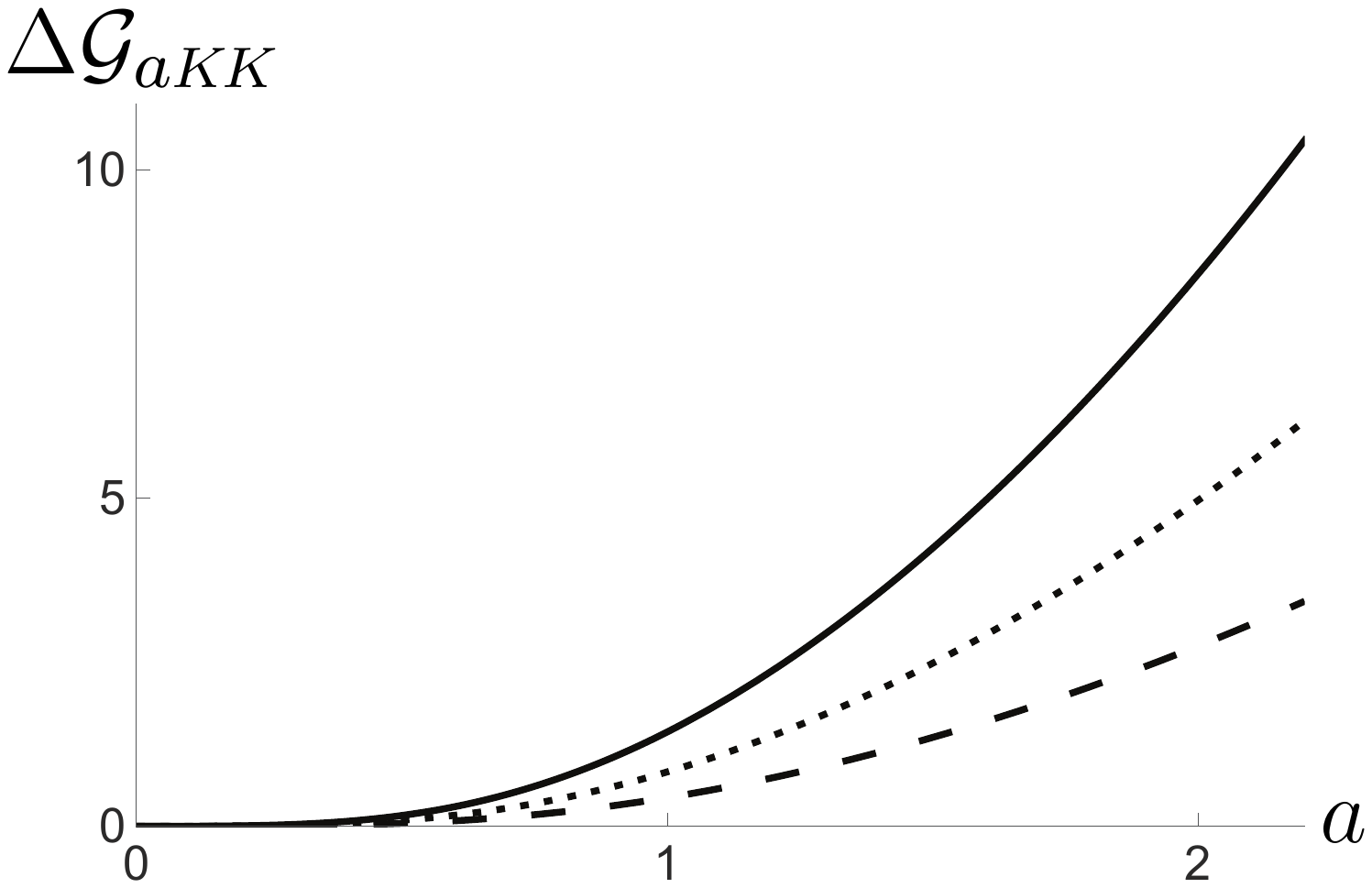} %
\vspace*{-10mm}
\caption{Plot of the dimensionless quantity $\Delta {\cal G}_{aK\!K} \equiv \dfrac{12\pi^{2}u_{\ast}^{2}}{u_{\Lambda}^{3}N^{2}L}\Delta G_{aK\!K}$ as a function of the dimensionless parameter $au_{\ast}$, in units where $u_{\ast}=1$. The solid, dotted, and dashed lines correspond to a large compactification radius case ($(u_{K\!K}/u_{\ast})^{4}=0.1$), an intermediate compactification radius case ($(u_{K\!K}/u_{\ast})^{4}=0.5$), and a small compactification radius case ($(u_{K\!K}/u_{\ast})^{4}=0.9$), respectively.}
\label{F14}
\end{figure}

The regularized holographic fidelity susceptibility in the reduced noncommutative Yang--Mills theory, denoted by $\Delta G_{aK\!K}$, is given by 
\begin{align}
\label{610}
\Delta G_{aK\!K} = \dfrac{N^{2}}{12\pi^{2}}u_{\Lambda}^{3}L^{2}(l_{K\!K}-l_{\rm K\!KC})\,,
\end{align}
where $l_{\rm K\!KC}$ represents the commutative limit of the length $l_{K\!K}$. The numerical dependence of  $\Delta G_{aK\!K}$ on the parameter $au_{\ast}$ is illustrated in Fig.\ref{F14}. 
In the reduced theory, the regularized holographic fidelity susceptibility increases monotonically with the noncommutativity parameter $a$, regardless of the magnitude of the parameter $u_{K\!K}$. Its sensitivity to $a$ diminishes as $u_{K\!K}$ increases, with the slope approaching zero in the limit $u_{K\!K} \to u_{\ast}$. Consistent with this, Fig.\ref{F14} shows that varying the compactification radius does not qualitatively modify the behavior of the regularized holographic fidelity susceptibility. 

%
%
\section{Conclusions}

In this paper, we compute the holographic subregion complexity (HSC) of a noncommutative deformation of Yang--Mills theory using the complexity=volume (CV) conjecture. Owing to the near-boundary behavior of the bulk metric, the HSC computed in the CV conjecture is ultraviolet divergent and therefore requires a UV cutoff. Nevertheless, we have shown that, in noncommutative Yang--Mills theory, the HSC we evaluated is cutoff-independent: the extracted contribution is universal, i.e., independent of the choice of regularization scheme.

Our results show that the HSC is sensitive to spatial noncommutativity. In noncommutative Yang--Mills theory, the characteristic length of the subregion $l$ is bounded below by a minimum length $l_{\rm min}$ set by the noncommutativity scale, $l_{\rm min} \sim a$. Consequently, when $l$ becomes comparable to $a$, the value of the HSC deviates from its commutative limit. Moreover, interpreting $u_{\ast}$ as the energy scale of the subregion, the HSC exhibits a sharp transition at $u_{\ast} \sim 1/a$; its behavior is qualitatively different in the low-energy $(u_{\ast} < 1/a)$ and high-energy $(u_{\ast} > 1/a)$ regimes.

We further computed the holographic fidelity susceptibility (HFS) in noncommutative Yang--Mills theory, following the conjecture proposed in Ref.~\cite{MAAFA}. The HFS exhibits qualitatively distinct behavior in the low-energy $(u_{\ast} < 1/a)$ and high-energy $(u_{\ast} > 1/a)$ regimes. Moreover, it increases monotonically with the noncommutativity parameter $a$; this monotonicity is insensitive to temperature and persists under dimensional reduction.

Furthermore, we examined the strong subadditivity property of the HSC and found that it is satisfied in both the commutative and noncommutative theories over a broad range of parameters. In the noncommutative case, however, as the overlap width $x$ approaches the minimum length $l_{\rm min}$, 
the result departs progressively from its commutative counterpart, and the strong subadditivity fails abruptly at $x=l_{\rm min}$. This breakdown induced by the minimum length is a distinctive feature of noncommutative theories and suggests that the non-local nature of noncommutative theories fundamentally alters the information-theoretic structure of the holographic dual.

At finite temperature, thermal effects substantially modify the HSC ${\cal C}_{AT}$ in the large characteristic length $l_{T}$ regime. In particular, its asymptotic dependence on $l_{T}$ crosses over to a volume-law-like behavior ${\cal C}_{AT} \propto l_{T}$. In addition, the HFS exhibits a sharp peak at $u_{\ast}=u_{T}$ in both the noncommutative theory and its commutative limit, indicative of phase-transition-like behavior. Another notable feature is that the temperature $u_{T}$ amplifies the response to noncommutativity: the slope of the regularized holographic fidelity susceptibility with respect to $a$ increases as the temperature rises.

Even after dimensional reduction, the HSC in the noncommutative theory changes sign near the minimum length scale $l = l_{\rm min}$. This behavior can be interpreted as a transition from a connected to a disconnected Ryu–-Takayanagi surface configuration. Consequently, this sign reversal may represent a kind of phase transition, if HSC can serve as an order parameter of a phase transition \cite{IRKDKAM}, although this interpretation warrants further investigation. Notably, no such phenomenon is observed in the commutative limit. By contrast, the HFS diverges as $u_{K\!K} \to 0$, indicating a non-analytic change associated with the shift in spatial dimensionality; this divergence occurs irrespective of whether the theory is noncommutative or commutative. In addition, in the reduced theory, the regularized holographic fidelity susceptibility becomes nearly insensitive to the noncommutativity parameter $a$ as  $u_{K\!K} \to 0$. This suggests that sensitivity to the noncommutative scale is lost when the subregion’s characteristic scale approaches the compactification scale.

In this work, we have investigated the HSC and HFS for a rectangular subregion. An important next step  is to determine whether the properties derived in this setup persist when the geometry of the subregion is modified. It is also crucial to assess the extent to which these properties depend on the dimensionality of the subregion. Complexity serves as an effective indicator for understanding physical phenomena such as phase transitions \cite{JYARF}. It would be particularly interesting to study phase transitions in noncommutative gauge theories through noncommutative deformations of complexity based on the CV conjecture. Moreover, since Krylov complexity is known to act as an order parameter in confinement--deconfinement transitions \cite{TANIMN}, it is important to investigate Krylov complexity in noncommutative gauge theories. Another direction for investigation is to examine whether the various properties derived here from the noncommutative deformation depend on the specific manner in which the deformation is implemented. For instance, it would be interesting to explore to what extent these properties are modified if the noncommutativity parameter is allowed to be a non-constant \cite{TIEJMSK, TIEJMSK2}. We hope to discuss this in the future.

\section*{Acknowledgments}

The author would like to thank S. Deguchi, E. Umezawa, Y. Ohtake and K. Suzuki for helpful discussions. This work was supported in part by the Dean's Grant for Specified Incentive Research, College of Engineering, Nihon University.

\clearpage 

%
%

\bibliographystyle{ytphys}
\bibliography{references_tn9}

\providecommand{\href}[2]{#2}\begingroup\raggedright\begin{thebibliography}{10}

\bibitem{LS2}
L.~Susskind, {\em {Three Lectures on Complexity and Black Holes}}.
\newblock Springer, 2020.
\newblock \href{http://arxiv.org/abs/1810.11563}{{\ttfamily arXiv:1810.11563
  [hep-th]}}.

\bibitem{MAN}
M.~A. Nielsen, ``{A geometric approach to quantum circuit lower bounds},'' {\em
  Quantum Info. Comput.} {\bfseries 6} (2006) 213,
  \href{http://arxiv.org/abs/quant-ph/0502070}{{\ttfamily
  arXiv:quant-ph/0502070}}.

\bibitem{ARBLS}
A.~R. Brown and L.~Susskind, ``{Second law of quantum complexity},''
  \href{http://dx.doi.org/10.1103/PhysRevD.97.086015}{{\em Phys. Rev. D}
  {\bfseries 97} (2018) 086015},
  \href{http://arxiv.org/abs/1701.01107}{{\ttfamily arXiv:1701.01107
  [hep-th]}}.

\bibitem{LS1}
L.~Susskind, ``{Computational Complexity and Black Hole Horizons},'' {\em
  Fortsch. Phys.} {\bfseries 64} (2016) 24--43,
  \href{http://arxiv.org/abs/1403.5695}{{\ttfamily arXiv:1403.5695 [hep-th]}}.
  [Addendum: Fortsch. Phys. 64, 44 (2016)].

\bibitem{DSLS1}
D.~Stanford and L.~Susskind, ``{Complexity and Shock Wave Geometries},'' {\em
  Phys. Rev. D} {\bfseries 90} (2014) 126007,
  \href{http://arxiv.org/abs/1406.2678}{{\ttfamily arXiv:1406.2678 [hep-th]}}.

\bibitem{ARBDARLSBSYZ1}
A.~R. Brown, D.~A. Roberts, L.~Susskind, B.~Swingle, and Y.~Zhao,
  ``{Holographic Complexity Equals Bulk Action?},'' {\em Phys. Rev. Lett.}
  {\bfseries 116} (2016) 191301,
  \href{http://arxiv.org/abs/1509.07876}{{\ttfamily arXiv:1509.07876
  [hep-th]}}.

\bibitem{ARBDARLSBSYZ2}
A.~R. Brown, D.~A. Roberts, L.~Susskind, B.~Swingle, and Y.~Zhao,
  ``{Complexity, action, and black holes},'' {\em Phys. Rev. D} {\bfseries 93}
  (2016) 086006, \href{http://arxiv.org/abs/1512.04993}{{\ttfamily
  arXiv:1512.04993 [hep-th]}}.

\bibitem{MA}
M.~Alishahiha, ``{Holographic Complexity},'' {\em Phys. Rev. D} {\bfseries 92}
  (2015) 126009, \href{http://arxiv.org/abs/1509.06614}{{\ttfamily
  arXiv:1509.06614 [hep-th]}}.

\bibitem{SRTT1}
S.~Ryu and T.~Takayanagi, ``{Holographic derivation of entanglement entropy
  from AdS/CFT},'' {\em Phys. Rev. Lett.} {\bfseries 96} (2006) 181602,
  \href{http://arxiv.org/abs/hep-th/0603001}{{\ttfamily arXiv:hep-th/0603001}}.

\bibitem{SRTT2}
S.~Ryu and T.~Takayanagi, ``{Aspects of Holographic Entanglement Entropy},''
  {\em JHEP} {\bfseries 08} (2006) 045,
  \href{http://arxiv.org/abs/hep-th/0605073}{{\ttfamily arXiv:hep-th/0605073}}.

\bibitem{OBADC}
O.~Ben-Ami and D.~Carmi, ``{On Volumes of Subregions in Holography and
  Complexity},'' {\em JHEP} {\bfseries 11} (2016) 129,
  \href{http://arxiv.org/abs/1609.02514}{{\ttfamily arXiv:1609.02514
  [hep-th]}}.

\bibitem{PRTS}
P.~Roy and T.~Sarkar, ``{Note on subregion holographic complexity and
  renormalization group flows},'' {\em Phys. Rev. D} {\bfseries 97} (2018)
  086018, \href{http://arxiv.org/abs/1708.05313}{{\ttfamily arXiv:1708.05313
  [hep-th]}}.

\bibitem{ABABSM}
A.~Banerjee, A.~Bhattacharya, and S.~Maulik, ``{HEE and HSC for flavors:
  perturbative structure in open string geometries},'' {\em JHEP} {\bfseries
  04} (2021) 212, \href{http://arxiv.org/abs/2008.02705}{{\ttfamily
  arXiv:2008.02705 [hep-th]}}.

\bibitem{SKJYZ}
S.-K. Jian and Y.~Zhang, ``{Subsystem Complexity and Measurements in
  Holography},'' {\em JHEP} {\bfseries 05} (2024) 241,
  \href{http://arxiv.org/abs/2312.04437}{{\ttfamily arXiv:2312.04437
  [hep-th]}}.

\bibitem{MMTNNSTTKW}
M.~Miyaji, T.~Numasawa, N.~Shiba, T.~Takayanagi, and K.~Watanabe, ``{Distance
  between Quantum States and Holographic Information Metric},'' {\em Phys. Rev.
  Lett.} {\bfseries 115} (2015) 261602,
  \href{http://arxiv.org/abs/1507.07555}{{\ttfamily arXiv:1507.07555
  [hep-th]}}.

\bibitem{ACMRDAS}
A.~Connes, M.~R. Douglas, and A.~Schwarz, ``{Noncommutative geometry and Matrix
  theory: Compactification on tori},'' {\em JHEP} {\bfseries 02} (1998) 003,
  \href{http://arxiv.org/abs/hep-th/9711162}{{\ttfamily arXiv:hep-th/9711162}}.

\bibitem{MRDCH}
M.~R. Douglas and C.~M. Hull, ``{D-branes and the noncommutative torus},'' {\em
  JHEP} {\bfseries 02} (1998) 008,
  \href{http://arxiv.org/abs/hep-th/9711165}{{\ttfamily arXiv:hep-th/9711165}}.

\bibitem{FAHASMMSJ}
F.~Ardalan, H.~Arfaei, and M.~M. Sheikh-Jabbari, ``{Noncommutative geometry
  from strings and branes},'' {\em JHEP} {\bfseries 02} (1999) 016,
  \href{http://arxiv.org/abs/hep-th/9810072}{{\ttfamily arXiv:hep-th/9810072}}.

\bibitem{NSEW}
N.~Seiberg and E.~Witten, ``{String theory and noncommutative geometry},'' {\em
  JHEP} {\bfseries 09} (1999) 032,
  \href{http://arxiv.org/abs/hep-th/9908142}{{\ttfamily arXiv:hep-th/9908142}}.

\bibitem{MRS}
S.~Minwalla, M.~V. Raamsdonk, and N.~Seiberg, ``{Noncommutative perturbative
  dynamics},'' {\em JHEP} {\bfseries 02} (2000) 020,
  \href{http://arxiv.org/abs/hep-th/9912072}{{\ttfamily arXiv:hep-th/9912072}}.

\bibitem{AA}
A.~Armoni, ``{Comments on perturbative dynamics of noncommutative Yang-Mills
  theory},'' {\em Nucl. Phys. B} {\bfseries 593} (2001) 229--242,
  \href{http://arxiv.org/abs/hep-th/0005208}{{\ttfamily arXiv:hep-th/0005208}}.

\bibitem{taka_naka-suzu}
H.~Takahashi, T.~Nakajima, and K.~Suzuki, ``{D1 / D5 system and Wilson loops in
  (non)commutative gauge theories},'' {\em Phys. Lett. B} {\bfseries 546}
  (2002) 273--281, \href{http://arxiv.org/abs/hep-th/0206081}{{\ttfamily
  arXiv:hep-th/0206081}}.

\bibitem{NST}
T.~Nakajima, K.~Suzuki, and H.~Takahashi, ``{Glueball mass spectra for
  supergravity duals of noncommutative gauge theories},'' {\em JHEP} {\bfseries
  01} (2006) 016, \href{http://arxiv.org/abs/hep-th/0508054}{{\ttfamily
  arXiv:hep-th/0508054}}.

\bibitem{TN_YO_KS}
T.~Nakajima, Y.~Ohtake, and K.~Suzuki, ``{The spectrum of low spin mesons at
  finite temperature in holographic noncommutative QCD},'' {\em Int. J. Mod.
  Phys. A} {\bfseries 28} (2013) 1350171,
  \href{http://arxiv.org/abs/1310.0393}{{\ttfamily arXiv:1310.0393 [hep-th]}}.

\bibitem{TN_YO_KS2}
T.~Nakajima, Y.~Ohtake, and K.~Suzuki, ``{Chiral Symmetry Restoration in
  Holographic Noncommutative QCD},'' {\em JHEP} {\bfseries 09} (2011) 054,
  \href{http://arxiv.org/abs/1011.2906}{{\ttfamily arXiv:1011.2906 [hep-th]}}.

\bibitem{TN_YO_KS3}
T.~Nakajima, Y.~Ohtake, and K.~Suzuki, ``{Baryon number current in holographic
  noncommutative QCD},'' {\em Phys. Rev. D} {\bfseries 96} (2017) 046018,
  \href{http://arxiv.org/abs/1702.06989}{{\ttfamily arXiv:1702.06989
  [hep-th]}}.

\bibitem{TN8}
T.~Nakajima, ``{Universal terms for holographic entanglement entropy in
  noncommutative Yang-Mills theory},'' {\em Phys. Rev. D} {\bfseries 103}
  (2021) 086005, \href{http://arxiv.org/abs/2006.14165}{{\ttfamily
  arXiv:2006.14165 [hep-th]}}.

\bibitem{NS_TT}
N.~Shiba and T.~Takayanagi, ``{Volume Law for the Entanglement Entropy in
  Non-local QFTs},'' {\em JHEP} {\bfseries 02} (2014) 033,
  \href{http://arxiv.org/abs/1311.1643}{{\ttfamily arXiv:1311.1643 [hep-th]}}.

\bibitem{JLKCR}
J.~L. Karczmarek and C.~Rabideau, ``{Holographic entanglement entropy in
  nonlocal theories},'' {\em JHEP} {\bfseries 10} (2013) 078,
  \href{http://arxiv.org/abs/1307.3517}{{\ttfamily arXiv:1307.3517 [hep-th]}}.

\bibitem{UKCNDSJSMW}
U.~Kol, C.~Nunez, D.~Schofield, J.~Sonnenschein, and M.~Warschawski,
  ``{Confinement, Phase Transitions and non-Locality in the Entanglement
  Entropy},'' {\em JHEP} {\bfseries 06} (2014) 005,
  \href{http://arxiv.org/abs/1403.2721}{{\ttfamily arXiv:1403.2721 [hep-th]}}.

\bibitem{DWP}
D.-W. Pang, ``{On holographic entanglement entropy of non-local field
  theories},'' {\em Phys. Rev. D} {\bfseries 89} (2014) 126005,
  \href{http://arxiv.org/abs/1404.5419}{{\ttfamily arXiv:1404.5419 [hep-th]}}.

\bibitem{LBCF}
J.~L.~F. Barbon and C.~A. Fuertes, ``{Holographic entanglement entropy probes
  (non)locality},'' {\em JHEP} {\bfseries 04} (2008) 096,
  \href{http://arxiv.org/abs/0803.1928}{{\ttfamily arXiv:0803.1928 [hep-th]}}.

\bibitem{WFAKSK}
W.~Fischler, A.~Kundu, and S.~Kundu, ``{Holographic Entanglement in a
  Noncommutative Gauge Theory},'' {\em JHEP} {\bfseries 01} (2014) 137,
  \href{http://arxiv.org/abs/1307.2932}{{\ttfamily arXiv:1307.2932 [hep-th]}}.

\bibitem{TJZX}
T.~Jia and Z.~Xu, ``{Noncommutativity and Holographic Entanglement Entropy},''
  {\em Phys. Rev. D} {\bfseries 95} (2017) 066002,
  \href{http://arxiv.org/abs/1612.04857}{{\ttfamily arXiv:1612.04857
  [hep-th]}}.

\bibitem{JCSEWFMLX}
J.~Couch, S.~Eccles, W.~Fischler, and M.-L. Xiao, ``{Holographic complexity and
  noncommutative gauge theory},'' {\em JHEP} {\bfseries 03} (2018) 108,
  \href{http://arxiv.org/abs/1710.07833}{{\ttfamily arXiv:1710.07833
  [hep-th]}}.

\bibitem{GKMBSPSRR}
G.~Katoch, M.~S. Balusu, S.~Parihar, and S.~R. Roy, ``{Quantum Complexity of
  Nonlocal Field Theories},'' {\em arXiv preprint arXiv:2511.00649} (2024) .
  (Check if published by 2026).

\bibitem{MAAFA}
M.~Alishahiha and A.~F. Astaneh, ``{Holographic Fidelity Susceptibility},''
  {\em Phys. Rev. D} {\bfseries 96} (2017) 086004,
  \href{http://arxiv.org/abs/1705.01834}{{\ttfamily arXiv:1705.01834
  [hep-th]}}.

\bibitem{HI}
A.~Hashimoto and N.~Itzhaki, ``{Noncommutative Yang-Mills and the AdS / CFT
  correspondence},'' {\em Phys. Lett. B} {\bfseries 465} (1999) 142--147,
  \href{http://arxiv.org/abs/hep-th/9907166}{{\ttfamily arXiv:hep-th/9907166}}.

\bibitem{MR}
J.~M. Maldacena and J.~G. Russo, ``{Large N limit of noncommutative gauge
  theories},'' {\em JHEP} {\bfseries 09} (1999) 025,
  \href{http://arxiv.org/abs/hep-th/9908134}{{\ttfamily arXiv:hep-th/9908134}}.

\bibitem{AOSJ}
M.~Alishahiha, Y.~Oz, and M.~M. Sheikh-Jabbari, ``{Supergravity and large N
  noncommutative field theories},'' {\em JHEP} {\bfseries 11} (1999) 007,
  \href{http://arxiv.org/abs/hep-th/9909215}{{\ttfamily arXiv:hep-th/9909215}}.

\bibitem{MLYSW}
M.~Li and Y.-S. Wu, ``{Holography and noncommutative Yang-Mills theory},'' {\em
  Phys. Rev. Lett.} {\bfseries 84} (2000) 2084--2087,
  \href{http://arxiv.org/abs/hep-th/9909085}{{\ttfamily arXiv:hep-th/9909085}}.

\bibitem{LAJM}
A.~Lewkowycz and J.~Maldacena, ``{Generalized gravitational entropy},'' {\em
  JHEP} {\bfseries 08} (2013) 090,
  \href{http://arxiv.org/abs/1304.4926}{{\ttfamily arXiv:1304.4926 [hep-th]}}.

\bibitem{TNTT}
T.~Nishioka and T.~Takayanagi, ``{AdS Bubbles, Entropy and Closed String
  Tachyons},'' {\em JHEP} {\bfseries 01} (2007) 090,
  \href{http://arxiv.org/abs/hep-th/0611035}{{\ttfamily arXiv:hep-th/0611035}}.

\bibitem{SJG}
S.-J. Gu, ``{Fidelity approach to quantum phase transitions},'' {\em Int. J.
  Mod. Phys. B} {\bfseries 24} (2010) 4371,
  \href{http://arxiv.org/abs/0811.3127}{{\ttfamily arXiv:0811.3127
  [quant-ph]}}.

\bibitem{KBDMMAA}
K.~Bamba, D.~Momeni, and M.~A. Ajmi, ``{Holographic Entanglement Entropy,
  Complexity, Fidelity Susceptibility and Hierarchical UV/IR Mixing Problem in
  AdS$_2$/open strings},'' {\em Int. J. Mod. Phys. A} {\bfseries 33} (2018)
  1850100, \href{http://arxiv.org/abs/1806.02209}{{\ttfamily arXiv:1806.02209
  [hep-th]}}.

\bibitem{MHTT}
M.~Headrick and T.~Takayanagi, ``{A Holographic proof of the strong
  subadditivity of entanglement entropy},'' {\em Phys. Rev. D} {\bfseries 76}
  (2007) 106013, \href{http://arxiv.org/abs/0704.3719}{{\ttfamily
  arXiv:0704.3719 [hep-th]}}.

\bibitem{TNSRTT}
T.~Nishioka, S.~Ryu, and T.~Takayanagi, ``{Holographic Entanglement Entropy: An
  Overview},'' {\em J. Phys. A} {\bfseries 42} (2009) 504008,
  \href{http://arxiv.org/abs/0905.0932}{{\ttfamily arXiv:0905.0932 [hep-th]}}.

\bibitem{GTHRCM}
G.~T. Horowitz and R.~C. Myers, ``{The AdS / CFT correspondence and a new
  positive energy conjecture for general relativity},'' {\em Phys. Rev. D}
  {\bfseries 59} (1998) 026005,
  \href{http://arxiv.org/abs/hep-th/9808079}{{\ttfamily arXiv:hep-th/9808079}}.

\bibitem{IRKDKAM}
I.~R. Klebanov, D.~Kutasov, and A.~Murugan, ``{Entanglement as a Probe of
  Confinement},'' {\em Nucl. Phys. B} {\bfseries 796} (2008) 274--293,
  \href{http://arxiv.org/abs/0709.2140}{{\ttfamily arXiv:0709.2140 [hep-th]}}.

\bibitem{TSSS1}
T.~Sakai and S.~Sugimoto, ``{Low energy properties of hadrons from holographic
  QCD},'' {\em Prog. Theor. Phys.} {\bfseries 113} (2005) 1083--1118,
  \href{http://arxiv.org/abs/hep-th/0412141}{{\ttfamily arXiv:hep-th/0412141}}.

\bibitem{TSSS2}
T.~Sakai and S.~Sugimoto, ``{More on holographic QCD},'' {\em Prog. Theor.
  Phys.} {\bfseries 114} (2005) 1083--1118,
  \href{http://arxiv.org/abs/hep-th/0507073}{{\ttfamily arXiv:hep-th/0507073}}.

\bibitem{JYARF}
J.~Yang and A.~R. Frey, ``{Complexity, scaling, and a phase transition},'' {\em
  JHEP} {\bfseries 09} (2023) 029,
  \href{http://arxiv.org/abs/2307.08229}{{\ttfamily arXiv:2307.08229
  [hep-th]}}.

\bibitem{TANIMN}
T.~Anegawa, N.~Iizuka, and M.~Nishida, ``{Krylov complexity as an order
  parameter for deconfinement phase transitions at large N},'' {\em JHEP}
  {\bfseries 04} (2024) 119, \href{http://arxiv.org/abs/2401.01501}{{\ttfamily
  arXiv:2401.01501 [hep-th]}}.

\bibitem{TIEJMSK}
T.~Araujo, I.~Bakhmatov, E.~{\'O}. Colg{\'a}in, J.~i.~Sakamoto, M.~M.
  Sheikh-Jabbari, and K.~Yoshida, ``{Yang-Baxter $\sigma$-models, conformal
  twists, and noncommutative Yang-Mills theory},'' {\em Phys. Rev. D}
  {\bfseries 95} (2017) 105006,
  \href{http://arxiv.org/abs/1702.02861}{{\ttfamily arXiv:1702.02861
  [hep-th]}}.

\bibitem{TIEJMSK2}
T.~Araujo, I.~Bakhmatov, E.~{\'O}. Colg{\'a}in, J.~i.~Sakamoto, M.~M.
  Sheikh-Jabbari, and K.~Yoshida, ``{Conformal Twists, Yang-Baxter
  $\sigma$-models \& Holographic Noncommutativity},'' {\em J. Phys. A}
  {\bfseries 51} (2018) 235401,
  \href{http://arxiv.org/abs/1705.02063}{{\ttfamily arXiv:1705.02063
  [hep-th]}}.

\end{thebibliography}\endgroup

\end{document}